\documentclass[aps,prd,twocolumn,amssymb,eqsecnum,showpacs,nofootinbib]{revtex4}

\usepackage{bm}
\usepackage{amsmath}
\usepackage{euscript}
\usepackage{graphicx}

\DeclareMathOperator{\sgn}{sgn}

\begin{document}

\title{Gravitational wave snapshots of generic extreme mass ratio inspirals}

\author{Steve Drasco}
\affiliation{Jet Propulsion Laboratory, California Institute of
Technology, Pasadena, CA 91109}
\affiliation{Center for Radiophysics and Space Research, Cornell
University, Ithaca, NY 14853}

\author{Scott A.\ Hughes}
\affiliation{Department of Physics and MIT Kavli Institute, MIT, 77
Massachusetts Ave., Cambridge, MA 02139}

\begin{abstract}
Using black hole perturbation theory, we calculate the gravitational
waves produced by test particles moving on bound geodesic orbits about
rotating black holes.  The orbits we consider are generic ---
simultaneously eccentric and inclined.  The waves can be described as
having radial, polar, and azimuthal ``voices'', each of which can be
made to dominate by varying eccentricity and inclination.  Although
each voice is generally apparent in the waveform, the radial voice is
prone to overpowering the others.  We also compute the radiative
fluxes of energy and axial angular momentum at infinity and through
the event horizon.  These fluxes, coupled to a prescription for the
radiative evolution of the Carter constant, will be used in future
work to adiabatically evolve through a sequence of generic orbits.
This will enable the calculation of inspiral waveforms that, while
lacking certain important features, will approximate those expected
from astrophysical extreme mass ratio captures sufficiently well to
aid development of measurement algorithms on a relatively short
timescale.
\end{abstract}
\pacs{04.70.-s, 97.60.Lf}
\maketitle

\section{Introduction}
\label{s:intro}

One of the leading problems in modern general relativity research is
the analysis of binary systems.  This problem is quite nasty in
general, especially when the members of the binary are strongly
interacting and gravitational-wave (GW) emission is significant.  It
simplifies considerably in the extreme mass ratio limit, in which one
member of the binary is much smaller than the other (taken to be a
black hole).  This is largely because the small mass ratio makes it
possible to bring perturbative techniques to bear.

This paper reports the development of a key piece of infrastructure
needed for an approximate approach to this problem: the ability to
accurately calculate the instantaneous GW emission from arbitary bound
geodesic orbits of rotating Kerr black holes.  Our work generalizes
previous calculations which specialized the orbit to be either
circular (of constant Boyer-Lindquist coordinate radius
{\cite{circular}}) or equatorial (confined to the black hole's plane of
reflection symmetry {\cite{equatorial}}).

By knowing the GWs, we also calculate the flux of energy $E$ and axial
angular momentum $L_z$ that the radiation carries to infinity and down
the black hole's event horizon.  Global conservation allows us to
evolve the energy and angular momentum associated with the orbit,
fixing the evolution of two of the three ``constants'' which
characterize black hole orbits (up to initial conditions).  The
remaining constant, $Q$, is known as the Carter constant\footnote{The
constant $K = Q + (L_z - aE)^2$ is often used as well.}
{\cite{Carter,MTW}}.  In the limit of zero black hole spin, the hole
is spherically symmetric, geodesic orbits conserve total angular
momentum $L$, and Carter's constant is given by $Q = L^2 - L_z^2$;
though not strictly accurate, the intuitive interpretation ``$Q = $
the rest of the angular momentum'' is useful even for non-zero black
hole spin.  It has recently become clear that, when the system evolves
``slowly'' (quantified below), it is possible to fix the evolution of
$Q$ using techniques not too different from those used to evolve $E$
and $L_z$ {\cite{Mino, Hughes et al, Drasco Flanagan
Hughes,sthn2005}}.

For such slowly evolving systems, we can then evolve {\it all} of
the orbital constants; by doing so, we can build reasonably realistic
inspirals corresponding to generic initial conditions, as well as the
associated waveforms.  The qualifier ``reasonably realistic'' refers
to the fact that this procedure neglects, by construction, the
influence of ``conservative'' pieces of the small body's self
interaction; only the ``dissipative'' influence can be analyzed in
this way.  As we discuss further below, neglect of the conservative
influence will limit the reliability of our inspirals; for the
applications we have in mind, this limitation is a
reasonable trade off in order to build waveforms on a rapid timescale.

We now briefly summarize our major motivations for analyzing this
problem, sketch the techniques that we have developed and our major
results, and discuss the utility and limitations of our results.

\subsection{Astrophysical context}
\label{s:context}

Though an idealization of a much more difficult general problem, the
extreme mass ratio limit in fact corresponds precisely to an important
subset of astrophysical binaries: stellar mass compact objects
captured onto highly relativistic orbits of massive black holes in
galaxy cores.  Such binaries are expected to be created by multibody
scattering processes in the stellar cluster that surrounds these
central black holes.

It has now become clear that the central bulges of essentially all
galaxies contain a massive black hole with properties strongly
correlated to that of the bulge {\cite{fm2000,getal2000}}.  It also
appears likely that {\it mass segregation} (the sinking of massive
stellar objects to the bottom of a gravitational potential due to
equipartition of kinetic energy {\cite{spitzer69}}) is likely to drive
the largest stellar mass compact objects into the vicinity of the
massive black hole, increasing the probability of forming a capture
binary --- see, for example, Ref.\ {\cite{meg2000}} for discussion
pointing to the possibility of a ``minicluster'' of stellar mass black
holes in the center of our galaxy, and Ref.\ {\cite{munoetal2005}} for
evidence of an overabundance of black holes in binaries in the
galaxy's inner parsec.  Mass segregation is also seen numerically in
galactic center N-body simulations by Freitag {\cite{freitag2001}}.
The abundance of galactic black holes plus mass segregation suggest
that the capture and GW driven inspiral of stellar mass compact
objects into massive black holes may be a relatively common phenomenon
in the universe.

The GWs produced by these {\it extreme mass ratio inspirals} (EMRIs)
are thus ideal targets for low-frequency GW detectors, particularly
the space-based {\it LISA} antenna, currently under development as a
joint NASA-ESA mission {\cite{LISA,lisaesa}}.  By combining
estimates for the rate at which such captures are likely to occur (per
galaxy) with the projected sensitivity of {\it LISA} (which determines
the distance to which sources can be measured, and hence sets the
number of galaxies that are relevant), one finds that dozens to
thousands of EMRI events may be measured over {\it LISA}'s mission
lifetime {\cite{Gair et al}}.

Because the small body only slightly perturbs the spacetime of the
large black hole, the GWs are largely set by --- and therefore encode
--- the nature of the black hole's spacetime.  A typical measured EMRI
event is likely to last for $\sim 10^4 - 10^5$ orbits; by coherently
tracking the waveform's phase over these orbits, {\it LISA} should
determine parameters characterizing the binary with great precision.
Black hole masses and spins should be measured with fraction of a
percent accuracy or better {\cite{Barack Cutler}}; it should even be possible
to check whether the spacetime has the ``shape'' (higher multipolar
structure) required by the no-hair theorems of general relativity
{\cite{ryan95,ryan97,ch2004}}.  EMRI measurements are expected to
provide a cornucopia of data of interest to astrophysicists and
general relativists.

This promising astrophysical scenario is a major motivator for much of
the effort in this problem (certainly for the present authors).
Reliable theoretical models of the inspiral waves will be needed both
in order to maximize the science return from {\it LISA} data, but also
to assure detection of these events.  Ideally, the final
science measurement of a detected signal will be performed with a
``measurement template'' --- a model waveform computed accurately
enough that it remains in phase with a fractional accuracy of roughly
$1/\rho$ (where $\rho$ is the measured signal-to-noise ratio) over the
entire inspiral.  The techniques which we describe here cannot
construct waveforms accurate enough for this task; as we describe in
Sec.\ {\ref{ss:a_and_l}}, measurement templates require a more
accurate and complete analysis of the small body's self-interaction
than our techniques encompass.

Our goal instead is to develop waveform models that are sufficiently
reliable that they may be used to develop data analysis techniques for
EMRI {\it detection}.  EMRI waveforms are characterized by 14
parameters\footnote{A useful counting of these parameters is as
follows: the 2 masses, the black hole spin ${\bf S}$, the initial
relative position ${\bf r}_0$, the initial relative velocity ${\bf
v}_0$, and the binary's position relative to the observer ${\bf R}$.
Each vector has 3 components, for a total of 14 parameters.  Including
the smaller body's spin ${\bf s}$ raises the count to 17; fortunately,
${\bf s}$ can be neglected for our purposes {\cite{Barack Cutler}}.}.
The number of measurement templates that would be needed to fully
cover the 14 dimensional manifold of waveforms would render any
search for these waves by this method infeasible {\cite{Gair
et al}}.  Detection will instead be done hierarchically using
(relatively) short segments of the EMRI signal {\cite{Gair et al}}.
Each short segment need only match coherently to a model for
about $10^3 - 10^4$ orbits.  The accuracy requirements on such
``detection templates'' are less stringent than those needed for
measurement templates.

\subsection{Sketch of our calculation}
\label{ss:sketch}

The key pieces of our formalism have been summarized in depth in
previous papers, particularly {\cite{circular}}; here, we provide a very
brief sketch largely to set the context for the following discussion.

We use the Teukolsky equation {\cite{Teukolsky}} to calculate the
influence of a perturbation to a black hole spacetime.  This equation
describes the evolution of a complex scalar $\psi_4$ which is
constructed from the Weyl (vacuum) curvature of the spacetime; it is
basically a wave equation for the Weyl curvature, linearized around
the Kerr background\footnote{Indeed, Michael Ryan has shown that the
Teukolsky equation can be derived from the ``Penrose equation''
{\cite{mpryan74}}, a nonlinear wave equation for the Riemann curvature
tensor that is constructed from the Bianchi identity; see
{\cite{penrose60}}.}, with a source.  Schematically, the Teukolsky
equation is of the form
\begin{equation}
{\cal D}^2\psi_4 = {\cal T}[\vec z(t)]\;.
\label{eq:teukschematic}
\end{equation}
The general form of the operator ${\cal D}^2$ and the source function
${\cal T}$ can be found in {\cite{Teukolsky}}; $\vec z(t)$ represents
the worldline of the orbiting body.  Coordinate time $t$ (which
amounts to time as measured by distant observers) is a particularly
convenient parameterization for our purposes.  Derived forms of Eq.\
(\ref{eq:teukschematic}) relevant to our analysis are given in Sec.\
{\ref{ss:tsn}}.

A key thing to note at this point is that the source depends on the
orbiting body's worldline ${\vec z}(t)$.  In order to construct ${\cal
T}$, we use the ``zeroth'' order worldline, neglecting radiation.
This worldline is built from geodesics of the background black hole
spacetime; as we discuss in Sec.\ {\ref{ss:a_and_l}}, we
incur an important cost due to this setup.  From this worldline, we
build the source function, and then compute $\psi_4$.  This complex
scalar completely encodes the GW flux to distant observers, and down
the black hole's event horizon {\cite{HH}} (equivalently, the
tidal interaction of the hole with the orbiting body {\cite{hartle}}).
All of the quantities which we use to describe GW emission and GW
induced orbit evolution are encoded in $\psi_4$.

We take advantage of the fact that the Teukolsky equation is separable
--- $\psi_4$ can be expanded into Fourier and spheroidal harmonics,
allowing it to be computed mode-by-mode.  Each mode is characterized
by a spherical-harmonic-like integer index $l$, and by three integer
indices ($m$, $k$, $n$) that label harmonics of the three fundamental
frequencies that describe the orbits.  Thanks to the linear nature of
the Teukolsky equation, the modes are independent of one another.
Codes that solve for $\psi_4$ in this manner are thus easily
parallelized --- many mode contributions can be computed separately
and independently.  Indeed, we have found that the code developed for
this work exhibits almost perfect $1/N$ computation-time scaling (where $N$ is
the number of processors) {\cite{Hughes et al}}.

We incur an important cost by expanding $\psi_4$ in modes: expanding
formally requires that we understand our source's behavior for all
time, $-\infty < t < \infty$, in order that the Fourier integral
exist.  In practice, this means that the orbit cannot evolve
``quickly'': we require the radiation to backreact {\it
adiabatically}, so that over a typical ``orbital time'' $T_{\rm orb}$
the change in any parameter that should be constant is much smaller
than the parameter itself (e.g., $T_{\rm orb}\dot E \ll E$).  In
principle, the adiabaticity requirement could be circumvented by
working in the time domain --- solving the wave equation for $\psi_4$
directly rather than expanding in modes.  Indeed, time domain codes
have proven superior to frequency domain codes for analyzing problems 
without sources (i.e. black hole ringdown), and are a more natural approach to evolving radiation
propagating in a black hole spacetime {\cite{td1,td2,td3,td4}}.  
Although new techniques may eventually change this story \cite{Sopuerta et al},
at present, frequency domain codes appear better suited to
handle the point-like sources appropriate to the EMRI problem.

The next natural step is to approximate the inspiral and associated
waveform from a sequence of geodesic orbits.  Beginning with some
starting orbit, we compute $\psi_4$.  From it, we extract a snapshot
waveform, which instantaneously approximates the true waveform, as
well as the rates of change of the orbital constants.  These rates of
change tell us how the trajectory evolves from the present orbit to
the next orbit in the sequence.  By repeating this process many times,
we can build an adiabatic waveform that usefully approximates the true
waveform.  We have not yet taken this step for generic orbits, though
most of the tools needed to do so are now in hand
{\cite{Mino, Hughes et al, Drasco Flanagan Hughes,sthn2005}}.

\begin{table}
\begin{ruledtabular}
\begin{tabular}{c|c|c|c|c}
Ref.                 & $a \ne 0$  &  eccentric orbits &
inclined orbits  & evolve orbits \\ \hline
{\cite{ckp}}         &            & \checkmark        &
$\cdots$             & \checkmark    \\ \hline
{\cite{Finn Thorne}}  & \checkmark &                   &
                 & \checkmark    \\ \hline
{\cite{Shibata circular, circular}} & \checkmark &         &
\checkmark       &               \\ \hline
{\cite{Shibata equatorial, equatorial}}   & \checkmark & \checkmark      &
       &               \\ \hline
{\cite{circularII}}      & \checkmark &                   &
\checkmark       & \checkmark    \\ \hline
Here                 & \checkmark & \checkmark        &
\checkmark       &            
\end{tabular}
\end{ruledtabular}
\caption{\label{t:history} A sketch of the history of this technique.
The magnitude of the massive black hole's spin angular momentum is
given by $aM$.  Due to spherical symmetry all geodesics of
Schwarzschild ($a=0$) black holes are planar, and can be considered
equatorial.}
\end{table}

Table {\ref{t:history}} summarizes the history of this approach.  To
date, the program has been completed only for non-spinning black holes
{\cite{ckp}}, circular-equatorial orbits \cite{Finn Thorne} circular orbits {\cite{circularII}}, and equatorial
orbits\footnote{Although Refs.~\cite{equatorial} and \cite{Shibata
equatorial} do not use their snapshot data to compute model inspirals,
it is a straightforward extension to do so.}  \cite{Shibata
equatorial, equatorial}.  For the astrophysical EMRI problem, these
are unrealistic restrictions.  Since extreme mass ratio binaries are
created through capture processes, we expect inclination to be
randomly distributed.  We also expect the eccentricity to be
substantial --- although radiative backreaction strongly reduces
eccentricity, EMRI events have such large initial eccentricities that
an estimated 50\% of all observable EMRIs will have eccentricities $e
> 0.2$ as they approach their last stable orbit {\cite{Barack
Cutler}}.  We must understand generic EMRIs in order to realize the
event rates predicted in Ref.\ {\cite{Gair et al}}.

\subsection{Applicability and limitations of our approach}
\label{ss:a_and_l}

The ``flux balancing-adiabatic evolution'' approach we advocate here is, as
emphasized above, an approximation to the evolution of extreme mass
ratio binaries.  A more rigorous approach, upon which most workers in
this field are focused, is based on computation of the {\it self
force} --- the small body's self interaction with its own
gravitational perturbation.  The gravitational self force is analogous
to the Abraham-Lorentz-Dirac electromagnetic self force
{\cite{ALD,Dirac}}.  Its complete general relativistic formulation was
worked out in detail by Mino, Sasaki, and Tanaka {\cite{MST}} and by
Quinn and Wald {\cite{QW}}; Poisson {\cite{ericlivrev}} provides a
very readable summary of this formalism.  Developing this approach to
the point that one can build a code around it to study the evolution
of generic orbits of Kerr black holes remains some time in the future;
however, efforts are intense and progress is rapid
{\cite{cqg_special}}.

At least heuristically, the self force can be broken into two pieces:
a ``dissipative'' force which carries energy and angular momentum away
from the binary (and thus drives the smaller body's inspiral), and a
``conservative'' force which does not.  These forces are computed
relative to geodesics of the {\it background} spacetime --- the
spacetime of the large black hole.  The conservative force tells us
that, even in the absence of radiation emission, the trajectory is
modified relative to the background geodesics --- the spacetime is
deformed from that of a black hole, so that black hole geodesics do
not precisely describe the motion of the small body.

Our approach totally neglects the conservative self force --- by
construction, we can only analyze dissipative effects.  This can be
seen in the schematic description of our calculation near Eq.\
(\ref{eq:teukschematic}) --- the worldline used in the source function
${\cal T}$ is built from geodesics of the {\it background} spacetime,
without incorporating the influence of the conservative force.  We
have argued previously {\cite{Hughes et al}} that this should be
adequate for scoping out EMRI waveforms and exploring issues related
to {\it LISA} data analysis.  Our argument was based in part on how
the phase error resulting from our approach scales with the mass ratio
$\mu/M$ [Eq.\ (1.2) of Ref.\ {\cite{Drasco Flanagan Hughes}}].  That
scaling was in turn based on the idea that (quoting from Ref.\
{\cite{Hughes et al}}) ``{\it Dissipative terms accumulate secularly;
conservative terms do not}''.  In other words, dissipative aspects of
the self force would accumulate phase effects over an EMRI event; we
expected that conservative terms would oscillate, and thus not
contribute as strongly over an inspiral.

To our chagrin, it has now been clearly demonstrated that our claim
that conservative effects do not accumulate secularly is wrong
(although the scaling of phase errors with mass ratio which we derived
from this is correct).  Using a compelling toy model to describe the
influence of the conservative self force, Pound, Poisson, and Nickel
{\cite{ppn05}} show that there is a component to the phase evolution
missed by a ``dissipative only'' evolution.  A simple way to
understand this additional component is as follows: The background
geodesics are characterized by oscillatory motion with three
frequencies, $\Omega_\phi$, $\Omega_\theta$, and $\Omega_r$, where
$\Omega_x$ describes oscillations (or orbits) associated with the
coordinate $x$.  A conservative self force changes the ``potentials''
that determine orbital motion, and thus modifies these three
frequencies:
\begin{eqnarray}
\Omega_\phi &\to& \Omega_\phi + \delta\Omega_\phi
\nonumber\\
\Omega_\theta &\to& \Omega_\theta + \delta\Omega_\theta
\nonumber\\
\Omega_r &\to& \Omega_r + \delta\Omega_r\;.
\label{eq:freqshift}
\end{eqnarray}
The shifts are smaller than their frequencies by (roughly) the mass
ratio: $\delta\Omega_x \sim (\mu/M)\Omega_x$, where $\mu$ is the mass
of the smaller body and $M$ is the mass of the black hole.

Some of the most interesting strong field features are due to beating
between these frequencies.  For example, periastron precession (well
known in the solar system due to planetary perihelion precession)
occurs at
\begin{equation}
\Omega_{\rm PP} = \Omega_\phi - \Omega_r\;.
\label{eq:peri_prec}
\end{equation}
Another effect is Lense-Thirring precession, the rotation of
the orbital plane due to frame dragging; it occurs at
\begin{equation}
\Omega_{\rm LT} = \Omega_\phi - \Omega_\theta\;.
\label{eq:lt_prec}
\end{equation}
The conservative self force shifts these precessions:
\begin{eqnarray}
\delta\Omega_{\rm PP} &=& \delta\Omega_\phi - \delta\Omega_r\;,
\nonumber\\
\delta\Omega_{\rm LT} &=& \delta\Omega_\phi - \delta\Omega_\theta\;.
\label{eq:precshift}
\end{eqnarray}
It is likely that $\delta\Omega_\phi \simeq \delta\Omega_\theta$; they
are presumably exactly equal for Schwarzschild black holes (spherical
symmetry).  However, $\delta\Omega_r$ is likely to differ quite a bit
from the other two frequency shifts, importantly modifying periastron
precession.  We thus expect that eccentric orbits in particular will
be impacted by the conservative self force.  This, indeed, is what is
found by Pound, Poisson, and Nickel.

Our inability to incorporate this conservative effect into our
analysis limits its utility.  We advocate our approach primarily to
begin exploring the space of EMRI waveforms in preparation for {\it
LISA}'s EMRI data analysis.  The best understanding at the moment,
after Ref.~\cite{ppn05}, is that the phase errors scale like
$(\mu/M)^0$ and $(v/c)^{-3}$; consequently, in the regime $v\sim c$
relevant to {\it LISA} the errors are formally of order unity.  It is
thus not yet clear if these errors are large enough to prevent
adiabatic waveforms being useful for search templates.  At any rate,
though adiabatic waveforms may miss an important contribution to the
phase evolution of extreme mass ratio binaries, they accurately
represent the spectral spread that can be expected.  Experience from
circular {\cite{circular,circularII}} and equatorial
{\cite{ckp,equatorial}} studies shows that waves from strong field
orbits radiate significant power into high harmonics of the orbital
frequencies.  The {\it LISA} datastream is expected to contain many
simultaneous confused sources --- certainly millions of white dwarf
binaries (whose radiation is essentially monochromatic), perhaps $\sim
10^3$ simultaneous EMRI sources, all lying under the signals from
massive coalescing cosmological black hole binaries.  Learning to
detangle these many signals will require models for the different
waves which accurately describe their time and frequency overlap.  By
modeling the frequency spread of EMRI waves with good accuracy (though
not necessarily the frequency evolution), adiabatic waveforms should
be a very useful tool.

{\it If} it turns out to be possible to separate conservative and
dissipative self force contributions, it should be easy to modify our
approach to include the leading conservative effects: we would simply
replace the geodesic worldlines presently used with worldlines that
are augmented by these conservative effects.  We would then use this augmented worldline
to construct the Teukolsky source function.  Doing so would allow us
to calculate an inspiral that accurately accounted for the leading
dissipative {\it and} conservative effects\footnote{One might label a
self force that allows this separation ``compassionately
conservative''.}.

It should be strongly emphasized that this is a rather big ``if'':
Because the gravitational self force is a gauge dependent quantity
(though its impact on binary orbits must be gauge independent), it is
far from clear that one can cleanly separate the ``conservative'' from
the ``dissipative'' influence.  It may be that the only way to
discriminate the two influences is to run the equations of motion
which the self force implies through a self-consistent wave-generation
calculation {\cite{ppn05}}, in which case there is no need to separate
these influences in the first place.

\subsection{Organization of this paper}
\label{ss:org}

We begin this paper by summarizing, in Sec.\ {\ref{s:Bound black hole
orbits}}, relevant properties of bound Kerr black hole orbits.  These
orbits determine, in turn, the properites of the Teukolsky equation's
source term.  Our goal is to present enough detail to make it clear
how we build this term.  Of particular importance is a frequency
domain description of functions that are built from these orbits
{\cite{Schmidt,Drasco Hughes}}.

We next (Sec.\ \ref{s:Perturbing a black hole with an orbiting test
mass}) summarize the relevant details of the black hole perturbation
formalism that we use.  We go into some detail in this section.  This
is in part to make the analysis self-contained, but also to correct
some small errors that have appeared in previous papers, notably Ref.\
{\cite{circular}} (hopefully without introducing too many new errors).
We first review in Sec.\ {\ref{ss:tsn}} how Teukolsky equation
solutions are built.  This is a somewhat subtle numerical problem ---
due to the long-rangedness of the separated Teukolsky equation's
radial potential, the amplitudes of ingoing ($\propto e^{-i\omega
r^*}$) and outgoing ($\propto e^{i\omega r^*}$) radiative solutions
grow at different rates.  [The quantity $r^*$ is the Kerr ``tortoise
coordinate''; cf.\ Eq.\ (\ref{eq:rstarofr}) below.]  One component of
the solution can easily swamp the other, causing numerical resolution
to be lost.  Sasaki and Nakamura {\cite{SN}} found an equation with
short-ranged potential whose solutions are related to those of the
Teukolsky equation by a simple differential transformation; the
combined Sasaki-Nakamura-Teukolsky formalism makes for a very
well-behaved numerical problem.  We next (Sec.\ \ref{ss:Bound test
mass sources}) discuss explicitly how the source function is built,
taking advantage of the results that we presented in Sec.\
{\ref{s:Bound black hole orbits}}, and then discuss some pratical
issues related to the numerics (Sec.\ \ref{ss:Numerical
considerations}).  We wrap up this section with a brief discussion of
how the gravitational waveforms and fluxes of ``conserved'' quantities
are extracted from these solutions.

Section \ref{s:Numerical algorithm} discusses several practical issues
related to numerics.  We first describe the algorithms used to
numerically represent the geodesics, a key element for the Teukolsky
source term, before describing issues related to computing each of the
modes from which we build $\psi_4$.  We then discuss in great depth
the algorithms we have developed to truncate our modal expansion.
Strictly speaking, the number of modes that should be used to build
$\psi_4$ is infinite; picking a finite value to truncate this
expansion that is ``large enough'' is somewhat subtle.  We describe a
scheme in Sec.\ \ref{ss:Truncation} that has proven to be robust
enough to work well for our present purposes.  There is clearly room
for improvement, however; we describe some ways in which this
procedure could be made better.  Finally, we conclude in Sec.\
\ref{ss:Validation} with detailed discussion of validation tests that
we made against previous results in the appropriate limits.  We find
that the code agrees perfectly with results from Ref.\
{\cite{circular}} when the eccentricity is zero\footnote{The
``generic'' code is a direct descendent of the ``circular'' code used
in Ref.\ {\cite{circular}}, so it is perhaps not too surprising that
there is agreement in the circular limit.  However, {\it many} details
of the generic code's inner workings are significantly different from
the circular incarnation, so this agreement is far from trivial.}, and
agrees quite well with the ``equatorial'' code of Glampedakis and
Kennefick in {\cite{equatorial}} in the limit of zero inclination.
Except for some (very small) down-horizon modes that do not
significantly impact the overall flux, we typically find agreement at
the level of $10^{-3}$ or better with Glampedakis and Kennefick.  This
is the level of accuracy that they cite for their code.

Our main results are presented in Sec.\ \ref{s:Results}.  The modal
decomposition allows us to break the waveform into ``voices'':
\begin{equation}
H \equiv h_+ - ih_\times = \sum_{kn} H_{kn} e^{-i\omega_{kn}(t - r^*)}\;,
\label{eq:voice_def}
\end{equation}
where $\omega_{kn} \equiv k\Omega_\theta + n\Omega_r$.  [A sum over
$l$ and $m$, including oscillations with frequency $m\Omega_\phi$, is
hidden in the definition of $H_{kn}$; see Eq.\ (\ref{eq:Hkn_def})
below.]  The ``polar voice'' is composed of the terms with $n = 0$ and
$k \ne 0$; The ``radial voice'' is composed of the terms with $k = 0$
and $n \ne 0$; The ``azimuthal voice'' is the term with $k = n = 0$,
and the ``mixed voice'' is composed of the terms with $k \ne 0$ and $n
\ne 0$.  A similar voice-by-voice labeling can be applied to the
fluxes.  We find that the importance of the various voices is fairly
simply controlled by the eccentricity and the inclination angle: Polar
voices become progressively more important as the purely polar orbit
(inclination $90^\circ$) is approached from either direction; radial 
voices rapidly become important as eccentricity, $e$, approaches unity,
typically dominating the waves' spectrum by $e \sim 0.3 - 0.5$.  We
speculate that this multivoice character may facilitate approximations
in the design of GW detection schemes, making it possible to detect
the most important voices of signals, rather than needing to detect
the (rather ornate) chorus of all voices together.

Section \ref{s:Summary and future work} presents our final
conclusions, and outlines future work to which we plan to apply this
formalism.  Chief among these future tasks will be augmenting this
approach with an adiabatic scheme to evolve the Carter constant
{\cite{Drasco Flanagan Hughes,sthn2005}}, and then using the complete
flux data to build model adiabatic inspirals and their associated
waveforms (as discussed in Sec.\ {\ref{ss:sketch}}).  We are
optimistic that this can be completed relatively quickly since there
appear to be no major hurdles or issues of principle that must be
overcome first.  Producing adiabatic waveforms for initial data
analysis algorithm development will then ``just'' be a matter of
finding sufficient CPU power.

\section{Bound black hole orbits}
\label{s:Bound black hole orbits}

In this section we review the geodesic motion of a non-spinning test
mass on a bound orbit of a Kerr black hole.  Kerr orbits are not a new
subject of investigation \cite{Carter, Wilkins}, but interest has been
renewed recently because of their relevance to EMRI GWs \cite{Schmidt,
Mino, Drasco Hughes}.  The main new development has been Mino's
exploitation of the fact that these orbits are fundamentally periodic
entities \cite{Mino}.  The utility of exploiting this property is
discussed in detail in Ref.~\cite{Drasco Hughes}, and details of its
application to EMRIs can be found in Sec.\ 3 of Ref.~\cite{Drasco
Flanagan Hughes}.  Here we will discuss the relation between an
orbit's geometry and its natural frequencies.  Using a special choice
of initial conditions (the fiducial geodesics from Ref.~\cite{Drasco
Flanagan Hughes}) we also derive a new system of equations for
efficient numerical evaluation of these orbits.

Kerr black holes are characterized by two parameters: the mass $M$ of
the black hole, and the magnitude $aM$ of its spin angular momentum,
with $0\le a \le M$.  Throughout this paper, we will use
Boyer-Lindquist \cite{Boyer Lindquist} coordinates
$(t,r,\theta,\phi)$, with units chosen so that $G=c=1$.  The line
element for the Kerr geometry is then given by \cite{MTW}
\begin{eqnarray} 
\label{Kerr}
ds^2_\text{Kerr} &=& 
- \left( 1-\frac{2Mr}{\Sigma} \right) ~dt^2
+ \frac{\Sigma}{\Delta}~dr^2
+ \Sigma~d\theta^2 \nonumber \\
&&+ \left( r^2+a^2 + \frac{2Ma^2r}{\Sigma}\sin^2\theta \right)\sin^2\theta~d\phi^2 \nonumber \\
&&- \frac{4Mar}{\Sigma}\sin^2\theta~dt~d\phi,
\end{eqnarray}
where 
\begin{align}
& \Sigma = r^2 + a^2\cos^2\theta, & \Delta = r^2 - 2Mr + a^2. &
\end{align}

Bound black hole orbits admit four constants of the motion which allow
us to rewrite the geodesic equations as a system of first order
differential equations.  Three of these constants are fairly
straightforward --- if the test particle has 4-momentum $\vec{p}$,
these constants are
\begin{align}
&\vec{p}\cdot\vec{p} = -\mu^2,&
&\vec{ \partial_t } \cdot \vec{p} = -E,&
&\vec{ \partial_\phi } \cdot \vec{p} = L_z,&
\end{align}
where $\mu$ is its rest mass, $E$ is the orbit's energy, and $L_z$ is
its axial angular momentum.  Carter discovered a fourth constant $Q$
which allows the motion to be completely described by a system of
first order equations \cite{Carter}.  As discussed in the
Introduction, it is often useful to think of the Carter constant as
representing the geodesic's non-axial angular momentum (a
correspondance which is exact for non-rotating black holes).

Carter's first order equations can be written in the following form
\begin{align}
&\left(\frac{dr}{d\lambda}\right)^2 = V_r(r), &
&\frac{dt}{d\lambda} = V_t(r,\theta),&
\nonumber \\
&\left(\frac{d\theta}{d\lambda}\right)^2 = V_\theta(\theta), &
&\frac{d\phi}{d\lambda} = V_\phi(r,\theta).&
 \label{geodesics}
\end{align}
The functions $V_t$, $V_r$, $V_\theta$, and $V_\phi$ are shown in
Appendix \ref{s:Geodesic details}.  The Mino time parameter $\lambda$
is related to the test mass' proper time $\tau$ by
\begin{equation}
\frac{d\tau}{d\lambda} = \Sigma.
\end{equation}
Mino time decouples the radial and polar equations of motion so that
$V_r=V_r(r)$ and $V_\theta = V_\theta(\theta)$.  This property
appears to have been recognized by Carter; however, Mino appears to be
the first to use $\lambda$ to fully exploit the periodic nature of Kerr
orbits \cite{Mino}.  Since the first order equations for $r$ and
$\theta$ are purely quadratic in their derivatives, $r(\lambda)$ and
$\theta(\lambda)$ are periodic functions for bound orbits.  The functions
$dt/d\lambda$ and $d\phi/d\lambda$ are then biperiodic functions in the 
sense defined by  Ref.~\cite{Drasco Hughes}.  As a result, 
after subtracting a term proportional to $\lambda$, the coordinates 
$t(\lambda)$ and $\phi(\lambda)$  can be represented with a two dimensional 
Fourier series.   These properties, discussed in more detail below, are extremely powerful.

\begin{figure*}
\includegraphics[width = 0.95\textwidth]{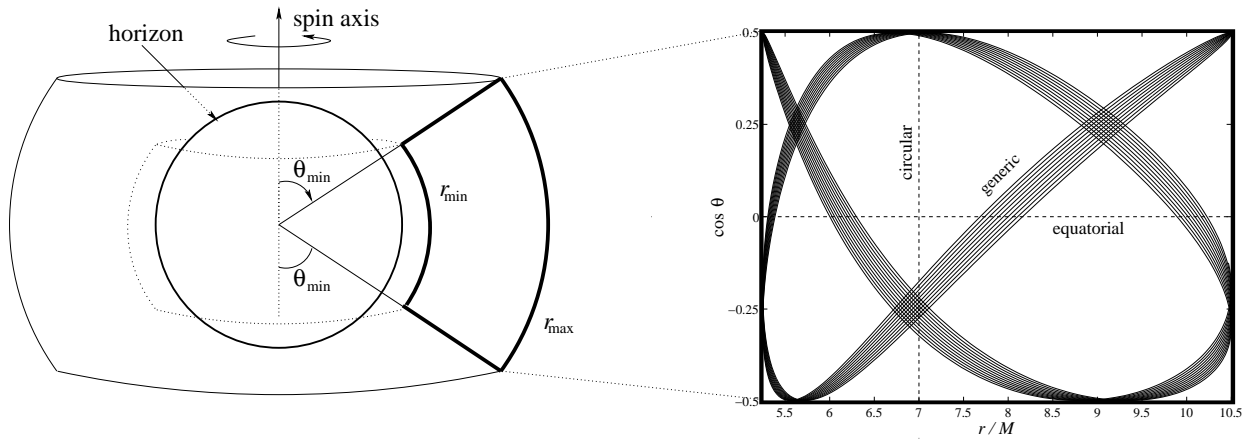}
\caption{The orbital torus and the evolution of $r$ and $\cos\theta$
for a generic geodesic orbit.  The magnitude of the black hole's spin
is $a = 0.998 M$.  The orbit shown here has
eccentricity $e = 1/3$, inclination $\theta_\text{inc} = 30^\circ$,
and semilatus rectum $p = 7$.  We start the orbit at $(r,\cos\theta) =
(r_{\min},\cos\theta_{\min})$, and end it around $(r,\cos\theta) =
(10.5M,-0.1)$, after several complete radial and polar
oscillations. The orbit is not closed: over time, it would eventually
fill the orbital torus.}
\label{f:torus-pretzle}
\end{figure*}

Solutions of the geodesic equations (\ref{geodesics}) are uniquely
determined if we specify $E$, $L_z$, $Q$ and the initial position
$\vec{z}(\lambda = 0)$ (or some other equivalent set of constants).  
We now focus on the orbit's geometry or
shape, which is determined by $E$, $L_z$, and $Q$.  This fact can be
roughly understood in that the orbit must be bounded by two radii
$r_{\min} \leq r \leq r_{\max}$, and [because the geometry
(\ref{Kerr}) is symmetric across the equatorial plane] one polar angle
$ \theta_{\min} \leq \theta \leq (\pi - \theta_{\min})$.  The orbit is
then confined to a toroidal region sketched in
Fig.~\ref{f:torus-pretzle}.
The boundaries of the orbital torus could have equivalently been
determined by an eccentricity $e$, a semilatus rectum $p$, and an
inclination $\theta_\text{inc}$ defined by
\begin{equation}
r_\text{min} = \frac{pM}{1+e},\;\;\;
r_\text{max} = \frac{pM}{1-e},\;\;\;
\theta_\text{inc} +(\sgn L_z) \theta_{\min} = \frac{\pi}{2}\;.
\end{equation}
Here we have included a factor of $\sgn L_z$ so as to make
$\theta_\text{inc}$ most closely resemble another common definition
for the orbital inclination angle $\iota$ \cite{circular}:
\begin{equation} \label{iota}
\cos \iota = \frac{L_z}{\sqrt{L_z^2 + Q}}.
\end{equation} 
A nice property of the angles $\theta_\text{inc}$ and $\iota$ (not 
shared by the angle $\theta_{\min}$) is that
they automatically encode a notion of prograde and retrograde ---
prograde orbits ($\phi$ motion parallel to the hole's rotation) have
$\theta_\text{inc},\iota < 90^\circ$; retrograde orbits ($\phi$ motion
antiparallel to the hole) have $\theta_\text{inc},\iota > 90^\circ$.
We have
found that in general $\iota \approx \theta_\text{inc}$. Even in the
strong field the two quantities typically differ by less than 10\%.
For the orbits which we discuss in detail here (Sec.~\ref{s:Results})
$\iota$ and $\theta_\text{inc}$ differ by no more than $0.5\%$.

Explicit algebraic relationships between the geometric orbital
parameters $(e,p,\theta_\text{inc})$ and the physical constants
$(E,L_z,Q)$ were first computed by Schmidt \cite{Schmidt}, and are
shown below in Appendix \ref{s:Geodesic details}.  We find that when
exploring the orbital parameters space, it is best to first think in
terms of $(e,p,\theta_\text{inc})$, and then to convert these
(following Appendix \ref{s:Geodesic details}) to $(E,L_z,Q)$ which are
used in all further calculations.

Using a special choice of initial conditions, we now derive a set of
equations for efficient numerical evaluation of bound orbits.  Since
the solutions to the radial and polar geodesic equations
(\ref{geodesics}) are periodic in Mino time, each can be expressed as
a Fourier series,
\begin{align} \label{Fseries}
&\theta(\lambda) = \sum_{k=-\infty}^\infty \theta_k e^{-ik\Upsilon_\theta \lambda}, &
&r(\lambda) = \sum_{n=-\infty}^\infty r_n e^{-in\Upsilon_r\lambda},&
\end{align}
where $\theta_k$ and $r_n$ are constants, and where
$\Upsilon_{r,\theta}$ are the orbital frequencies in Mino time.  In
Appendix \ref{s:Geodesic details} we list explicit expressions for
these frequencies as functions of the orbital parameters described in
the previous section.  It is convenient to write these series in terms
of angle-variables,
\begin{align}
&w_r = \Upsilon_r \lambda,&
&w_\theta = \Upsilon_\theta \lambda.&
\end{align}
The result is
\begin{align} \label{Fseries w}
&\theta(w_\theta) = \sum_{k=-\infty}^\infty \theta_k e^{-ik w_\theta}, &
&r(w_r) = \sum_{n=-\infty}^\infty r_n e^{-in w_r}.&
\end{align}

We now specialize to a specific choice of initial conditions for the
radial and polar motion.  We require that the orbit begins at a
turning point of both the $r$ and $\theta$ coordinates,
\begin{align} \label{initial rt}
&r(w_r=0) = r_{\min},&
&\theta(w_\theta=0) = \theta_{\min}.&
\end{align}
This choice of initial conditions will result in a simplified
numerical evaluation of the geodesics.  For example, the coordinates
$r$ and $\theta$ are now even functions of $w_{r,\theta}$,
\begin{align} \label{even coords}
&r(-w_r) = r(w_r),&
&\theta(-w_\theta) = r(w_\theta),&
\end{align}
and the Fourier series (\ref{Fseries}) become cosine series,
\begin{eqnarray}
\theta(w_\theta) &=& \theta_0 + 2\sum_{k = 1}^\infty \theta_k \cos(k w_\theta),
\\
r(w_r) &=& r_0 + 2\sum_{n = 1}^\infty r_n \cos(n w_r).
\end{eqnarray}
In this paper we will not evaluate these series
explicitly\footnote{Current investigations suggest that doing so may
lead to greater computational efficiency.}, though we will use the
fact that they are even functions.

We now derive a similarly simplified series expansion for $t$ and
$\phi$.  Both the $t$ and $\phi$ coordinates will be treated in a
similar way, so to save space we define
\begin{align}
&x = t,\phi,&
&\dot x = V_t,V_\phi,&
\end{align}
such that an overdot represents a derivative with respect to Mino time
$\lambda$.  Following the analysis preceeding Eqs.~(3.18) and (3.19)
in Ref.~\cite{Drasco Hughes}, we write the derivatives $\dot x$ in
the form
\begin{subequations}\label{dot series}
\begin{eqnarray} 
\dot x(\lambda) &=& \dot x_{00} + \sum_{k=1}^\infty (\dot x_k^\theta e^{-ik\Upsilon_\theta\lambda} + \text{c.c.})
\nonumber \\ 
	    && + \sum_{n=1}^\infty (\dot x_n^r  e^{-in \Upsilon_r\lambda} + \text{c.c.}),
\\
\dot x_k^\theta &=& \frac{1}{2\pi} \int_0^{2\pi} dw_\theta~ \dot x^\theta[\theta(w_\theta)] e^{i k w_\theta},
\\
\dot x_n^r &=& \frac{1}{2\pi} \int_0^{2\pi} dw_r~ \dot x^r[r(w_r)] e^{i n w_r}.
\end{eqnarray}
\end{subequations}
Here ``c.c.'' means the complex conjugate of the preceeding term.  The
leading constants are defined by
\begin{equation}
\dot x_{00} = \dot x_0^\theta + \dot x_0^r\;.
\end{equation}
We have made use of the fact that the derivatives of both $t$ and
$\phi$ separate into a sum of two functions, each depending only on
one of the coordinates\footnote{It's interesting to note that
Teukolsky has shown that the master equation (\ref{master}) separates
in any coordinates where this property (\ref{split potentials}) is
preserved \cite{Teukolsky}.},
\begin{equation} \label{split potentials}
{\dot x} = {\dot x}^\theta(\theta)+ {\dot x}^r(r).
\end{equation}
Following the notation of Ref.~\cite{Drasco Hughes}, we use the
following symbols for ${\dot x}_{00}$ in the cases $x=t$ and $x=\phi$:
\begin{align}
&\Gamma = (V_t)_{00},&
&\Upsilon_{\phi} = (V_\phi)_{00}.&
\end{align}
The constant $\Gamma$ is the analog of the Lorentz factor of special
relativity.  For example, in the case of an orbit which is both
circular and equatorial, we have $\Gamma = dt/d\lambda = (dt/d\tau)
\Sigma$ ($\Sigma$ is constant for circular-equatorial orbits).
$\Gamma$ also relates the Mino time frequencies
$\Upsilon_{\phi,\theta,r}$ to coordinate time frequencies $\Omega_r$,
$\Omega_\theta$, and $\Omega_\phi$:
\begin{align} \label{t frequencies}
&\Omega_\phi = \frac{\Upsilon_\phi}{\Gamma},&
&\Omega_\theta = \frac{\Upsilon_\theta}{\Gamma},&
&\Omega_r = \frac{\Upsilon_r}{\Gamma}.&
\end{align}
Schmidt \cite{Schmidt} provides an elegant derivation of closed form
expressions for the orbital frequencies.  His results were converted
into the Mino time frequencies in Ref.~\cite{Drasco Hughes} (see
Appendix \ref{s:Geodesic details} for details).

As a consequence of their initial values (\ref{initial rt}), the
functions $\theta(w_\theta)$ and $r(w_r)$ are even.  The functions
$\dot x^\theta(w_\theta)$ and $\dot x^r(w_r)$ are then also even, and
the series (\ref{dot series}) above simplifies to
\begin{subequations}\label{simple dot series}
\begin{eqnarray} 
\dot x(\lambda) &=& \dot x_{00} + 2\sum_{k=1}^\infty \dot x_k^\theta \cos(k\Upsilon_\theta\lambda)
 + 2\sum_{n=1}^\infty \dot x_n^r  \cos(n \Upsilon_r\lambda). \label{expanded xdot}
\nonumber \\ && \\
\dot x_k^\theta &=& \frac{1}{\pi} \int_0^{\pi} dw_\theta~ \dot x^\theta[\theta(w_\theta)] \cos(k w_\theta),
\\
\dot x_n^r &=& \frac{1}{\pi} \int_0^{\pi} dw_r~ \dot x^r[r(w_r)] \cos(n w_r).
\end{eqnarray}
\end{subequations}

If we now assume initial values for $t$ and $\phi$,
\begin{align}
&t(\lambda = 0) = 0\;,&
&\phi(\lambda = 0) = 0\;,&
\end{align}
so that we are now discussing the fiducial geodesics from
Ref.~\cite{Drasco Flanagan Hughes}, a similarly simplified expression
for the coordinates $x$ can be found by integrating the series
(\ref{expanded xdot}) for their derivatives $\dot x$:
\begin{subequations} 
\begin{align} \label{expanded x}
x(\lambda) &= {\dot x}_{00} \lambda + \Delta x^\theta(\lambda) + \Delta x^r(\lambda)&
\\
\Delta x^\theta(\lambda) &= \sum_{k=1}^{\infty} \Delta x_k^\theta \sin(k \Upsilon_\theta \lambda), &
\label{Dxtheta}
\\
\Delta x^r(\lambda) &= \sum_{n=1}^{\infty} \Delta x_n^r \sin(n \Upsilon_r \lambda), &
\label{Dxr}
\\
\Delta x_k^\theta &= \frac{2}{k\pi\Upsilon_k} \int_0^\pi dw_\theta~ {\dot x}^\theta(w_\theta) \cos(k w_\theta),&
\label{Dxk}
\\
\Delta x_n^r &= \frac{2}{n\pi\Upsilon_r} \int_0^\pi dw_r~ {\dot x}^r(w_r) \cos(n w_r).&
\label{Dxn}
\end{align}    
\end{subequations}
Note that in the cases of orbits with special geometries, the
coordinates $x$ simplify accordingly:
\begin{subequations}\label{geodesic sym}
\begin{align}
\Delta x^r(\lambda)      &= 0 = \Delta x^r_n      & &\text{(for circular orbits)}\;,  & \\
\Delta x^\theta(\lambda) &= 0 = \Delta x^\theta_k & &\text{(for equatorial orbits)}\;.&
\end{align}
\end{subequations}


Evaluating the integrands in Eqs.\ (\ref{Dxk}) and (\ref{Dxn}) as
functions of $r(w_r)$ and $\theta(w_\theta)$, would first require a
direct integration of the radial and polar geodesic equations.
Unfortunately this is somewhat difficult because the derivatives $V_r$
and $V_\theta$ vanish (by definition) at the orbital turning points.
This means, for example, that the integrand in the direct expression
for $r(\lambda)$,
\begin{equation}
\lambda = \int_{r_{\min}}^{r(\lambda)} \frac{dr'}{\pm \sqrt{V_r(r')}},
\end{equation} 
contains singularities which, although integrable, complicate the
numerics.  This is a well known problem (see the Appendix of
Ref.~\cite{Drasco Hughes} for more details) which can be avoided if we
change the integration variables to more well behaved coordinates
$\chi(\theta)$ and $\psi(r)$, where
\begin{align} \label{new coords}
&\cos\chi = \frac{\cos\theta}{\cos\theta_{\min}},&
&r = \frac{pM}{1+e\cos\psi}.&
\end{align}
Recall that $p$ is the semilatus rectum of the orbit, and $e$ is the
orbital eccentricity.  So in practice we replace Eqs.(\ref{Dxk}) and
(\ref{Dxn}) with
\begin{eqnarray} \label{final delta}
\Delta x_k^\theta &=& \frac{2}{k\pi\Upsilon_\theta} \int_0^\pi d\chi~\frac{dw_\theta}{d\chi} {\dot x}^\theta(w_\theta) \cos(k w_\theta),
\label{Dxk2}
\\
\Delta x_n^r &=& \frac{2}{n\pi\Upsilon_r} \int_0^\pi d\psi~ \frac{dw_r}{d\psi}{\dot x}^r(w_r) \cos(n w_r).
\label{Dxn2}
\end{eqnarray}
In order to use this result, we of course need explicit expressions
for $w_\theta(\chi)$, $w_r(\psi)$, and their derivatives.  These
functions were first derived in Ref.~\cite{Drasco Hughes}, and the
results are shown below in Appendix \ref{s:Geodesic details}.  It
turns out that $w_\theta(\chi)$ and $dw_\theta/d\chi$ can be written
in terms of elliptic integrals.  The functions $w_r(\psi)$ and
$dw_r/d\psi$ are slightly more involved.

We have found that this technique is an extremely efficient way to
evaluate bound orbits and functions of them (see Ref.~\cite{Drasco
Hughes} for a simple example).  The two expansions given by
Eqs.~(\ref{Dxtheta}) and (\ref{Dxr}) converge very rapidly.  Typically
only about $50$ terms are needed in order to obtain a fractional
accuracy of $10^{-12}$.

\section{Perturbing a black hole with an orbiting test mass}
\label{s:Perturbing a black hole with an orbiting test mass}

In this section we will discuss how a non-spinning test mass $\mu$ on
a bound geodesic perturbs the Kerr geometry (\ref{Kerr}).  The
radiative information describing the linear order perturbation can be
extracted from the Weyl curvature scalar \cite{Wald}
\begin{equation} \label{Weyl}
\psi_4 = -C_{\alpha\beta\gamma\delta} n^\alpha \bar m^\beta n^\gamma
 \bar m^\delta\;.
\end{equation}
Here, overbar denotes complex conjugation, $C_{abcd}$ is the Weyl
curvature tensor (the Riemann tensor in vacuum), and the vectors are
elements of a Newman-Penrose null-basis \cite{NP, Chandra} \{$
l^\alpha,m^\alpha,\bar m^\alpha,n^\alpha$\}.  For distant observers,
the curvature scalar $\psi_4$ is simply related to the metric
perturbation.  For these same observers, and also for observers at the
event horizon, $\psi_4$ is simply related to the fluxes of energy and
angular momentum (see Sec.~\ref{ss:Waveforms and fluxes} for details).
Since we are only analyzing perturbations to first order in the mass
ratio, $\psi_4$ always represents the leading $O(\mu/M)$ contribution
to the curvature perturbation.

\subsection{Teukolsky-Sasaki-Nakamura formalism}
\label{ss:tsn}

Teukolsky showed that $\psi_4$ is a solution to an equation of the 
form \cite{Teukolsky}
\begin{align} \label{master}
\left[ \widehat{U}_{t\phi r}(r) + \widehat{V}_{t\phi \theta}(\theta) \right] 
\varphi = -\EuScript{T},
\end{align}
where $\EuScript{T}$ is the source term (described below 
in Sec.~\ref{ss:Bound test mass sources}), 
\begin{align}
&\varphi = \rho^{-4}\psi_4, &
&\rho = -(r - ia\cos \theta)^{-1},&
\end{align}
and where $\widehat{U}_{t\phi r}(r)$ and $\widehat{V}_{t\phi
\theta}(\theta)$ are second order differential operators containing
derivatives with respect to the variables shown as subscripts [see
Eqs.~(\ref{Utphir}) and (\ref{Vtphitheta})].  Equation (\ref{master})
is known as the master equation.  In the next two sections, we will
summarize Teukolsky's technique for solving the master equation by
separation of variables \cite{Teukolsky}, converting
Eq.~(\ref{master}) into a pair of ordinary differential equations.
This technique is referred to as solving the master equation ``in the
frequency domain''.  As briefly discussed in the Introduction, one
could instead solve the master equation numerically as a partial
differential equation ``in the time domain'' (see Refs.~\cite{Poisson
PDE,Khanna} and references therein).  Such an approach can be
advantageous for problems where there is no source, or where the
source is not pointlike.  To date at least, time domain treatments are
accurate to about 10\% \cite{Khanna} for EMRI sources (at least for
quantities like orbit averaged fluxes of energy and angular momentum).

In the source free case $\EuScript{T} = 0$, the master equation
(\ref{master}) is satisfied by functions of the form
\begin{equation}
\varphi_m(\omega,C) = 
\widetilde{R}_m(r;\omega,C)S_m(\theta;\omega,C)e^{-i\omega t + im\phi},
\end{equation}
where $C$ is some constant, $m$ is an integer (since $\psi$ must be
periodic in $\phi$), and $\omega$ is any real number.  The functions
$\widetilde{R}$ and $S$ are solutions to ordinary differential
equations of the form
\begin{subequations}
\begin{align}
\left[ \widehat{U}_r(r,m,\omega) - C\right]\widetilde{R}_m(r,\omega,C)
&= 0,& \label{radial}
\\
\left[ \widehat{V}_\theta(\theta,m,\omega) + C\right]
S_m(\theta,\omega,C) &= 0.& \label{angular}
\end{align}
\end{subequations}
Requiring that the solutions to the angular equation (\ref{angular}) 
(known as spin-weighted spheroidal harmonics with spin weight $-2$) 
be regular results in a discrete spectrum of eigenvalues
$C = C_{lm}(\omega)$,
\begin{equation}
S_m[\theta,\omega,C_{lm}(\omega)] \equiv S_{lm}(\theta,\omega)\;,
\end{equation} 
where $l \ge \max(|m|,2)$ is an integer.
The functions $S_{lm}(\theta,\omega)$ are uniquely defined only after we specify 
boundary conditions, or equivalently  after we chose a normalization convention.    
We use the convention\footnote{Although not stated there, this is the normalization 
convention used in
Ref.~\cite{circular}.}
\begin{equation} \label{S norm}
\int_0^{\pi} d\theta~ \left[S_{lm}(\theta,\omega)\right]^2 \sin\theta
= \frac{1}{2\pi}\;.
\end{equation}
The choice of normalization is arbitrary in the sense that it does
not change any physical predictions.  However, in order to 
compare equations from other papers in which a different convention was
used, we must state our choice explicitly.
The functions $S_{lm}(\theta,\omega)$ will be used as a basis to
express functions $f(t,\theta,\phi)$ as follows:
\begin{equation}
f(t,\theta,\phi) = \sum_{l=2}^\infty \sum_{m=-l}^l
\int_{-\infty}^\infty\!\! d\omega\,\langle lm\omega|f\rangle
S_{lm}(\theta,\omega) e^{-i\omega t + i m \phi},
\end{equation}
where 
\begin{equation}
\left<lm\omega|f\right> = \frac{1}{2\pi}\int dt~ \int d\Omega~
f(t,\theta,\phi) S_{lm}(\theta,\omega) e^{i\omega t - i m\phi},
\end{equation}
so that we have $\left<l'm'\omega'|lm\omega\right> =
\delta_{ll'}\delta_{mm'}\delta(\omega - \omega')$.

We now return to the case where the source $\EuScript{T}$ in 
the master equation (\ref{master}) is nonvanishing.  Assuming a 
solution of the form
\begin{equation} \label{psi_4}
\varphi = \rho^{-4}\psi_4 = \sum_{l=2}^\infty \sum_{m=-l}^l \int d\omega~ \varphi_{lm}(\omega)\;,
\end{equation}
where
\begin{equation} \label{psilmomega}
\varphi_{lm}(\omega) =
R_{lm}(r,\omega)S_{lm}(\theta,\omega)e^{-i\omega t + im\phi},
\end{equation}
implies that the radial function $R_{lm}(r,\omega)$ must satisfy
\begin{equation} \label{Teukolsky}
\left[\Delta^2 \frac{d}{dr}\left( \frac{1}{\Delta} \frac{d}{dr}\right)
- \mathcal{V}_{lm}(r,\omega)\right] R_{lm}(r,\omega) = -
\EuScript{T}_{lm}(r,\omega),
\end{equation}
where $\EuScript{T}_{lm}(r,\omega) =
\left<lm\omega|\EuScript{T}\right>$, and the potential is given by
\begin{equation}
\mathcal{V}_{lm}(r,\omega) = -\frac{K^2+4i(r-M)K}{\Delta} + 8i\omega r + \lambda_{lm}(\omega).
\end{equation}
Here $K = (r^2 + a^2)\omega - ma$, and 
\begin{equation} \label{lambda eig}
\lambda_{lm}(\omega) = -C_{lm}(\omega) - 2am\omega.
\end{equation}
The radial equation (\ref{radial}) is just the source free
(homogeneous) version of the previous radial equation (\ref{Teukolsky}).  The
two independent solutions to the homogeneous radial equation will be
used to construct the solution to the inhomogeneous equation.  We use
the so-called ``in-up basis'' $\{ R^\text{H}_{lm}(r,\omega),
R^\infty_{lm}(r,\omega)\}$, which is defined by the following limiting
behavior \cite{Drasco Flanagan Hughes}:
\begin{subequations}\label{inup}
\begin{align}
R_{lm}^\text{H}(r\to r_+,\omega) &= B_{lm}^{\rm hole}(\omega)\Delta^2
e^{-i P r^*},&
\label{RH H} \\
R_{lm}^\text{H}(r\to \infty,\omega) &= B_{lm}^{\rm out}(\omega)r^3
  e^{i\omega r^*} + \frac{B_{lm}^{\rm in}(\omega)}{r} e^{-i\omega
  r^*},&
\label{RH inf}\\
R_{lm}^\infty(r\to r_+,\omega) &= D_{lm}^{\rm out}(\omega)e^{i P r^*}
  + D_{lm}^{\rm in}(\omega) \Delta^2 e^{-i P r^*},&
\label{Rinf H} \\
R_{lm}^\infty(r\to \infty,\omega) &= D_{lm}^{\infty}(\omega)r^3
e^{i\omega r^*}.&
\label{Rinf inf}
\end{align}
\end{subequations}
Here $P = \omega - ma/(2Mr_+)$, and the ``tortoise coordinate'' $r^*$
satisfies $dr^*/dr = (r^2 + a^2)/\Delta$:
\begin{equation}
r^*(r) = r + \frac{2M r_+}{r_+ - r_-}\ln\frac{r-r_+}{2 M}
 - \frac{2M r_-}{r_+ - r_-}\ln\frac{r-r_-}{2 M} ,
\label{eq:rstarofr}
\end{equation}
where $r_\pm = M \pm \sqrt{M^2 - a^2}$ are the roots of $\Delta$
($r=r_+$ is the location of the event horizon).

The numerical calculation of the homogeneous solutions, say
$R_{lm}^\text{H}(r)$, would in principle work as follows.  First set
$B_{lm}^\text{in}(\omega) = 1$ (a normalization convention).  Then
note that
\begin{align}
\frac{R_{lm}^\text{H}(r\to r_+)}{B_{lm}^{\rm hole}(\omega)} &=
\Delta^2 e^{-i P r^*},&
\\
\frac{R_{lm}^\text{H}(r\to \infty)}{B_{lm}^{\rm hole}(\omega)}
&= \frac{B_{lm}^{\rm out}(\omega)}{B_{lm}^{\rm hole}(\omega)}r^3
e^{i\omega r^*} + \frac{1}{B_{lm}^{\rm hole}(\omega)r} e^{-i\omega
r^*}.&
\label{inf ratio}
\end{align}
Next starting near the horizon at $r = r_+$, integrate forward.  When
we reach $r \approx \infty$, we can read off
$B^\text{out}/B^\text{hole}$ and $1/B^{\rm hole}$.  (When clarity
permits, we will often drop the somewhat cumbersome subscripts $l,m$
and the dependence on $r,\omega$.)  Since the first term in Eq.\
(\ref{inf ratio}) grows rapidly with $r$, the two terms cannot be
extracted with equal accuracy.  We work around this by instead solving
the homogeneous Sasaki-Nakamura equation \cite{SN} (which was designed
to avoid this problem) and converting the result to $R^{H,\infty}$;
see Ref.~\cite{circular} for details.  Recently Fujita and Tagoshi
have worked out a sophisticated numerical scheme for computing the
functions $R^{H,\infty}$ numerically with great accuracy (with
fractional accuracies $\sim10^{-14}$ in the corresponding energy
fluxes for orbits which are both circular and equatorial) \cite{Fujita
Tagoshi}.

The general solution to the radial equation (\ref{Teukolsky}),
corresponding to the retarded solution of the master equation
(\ref{master}), is
\begin{align}
R_{lm}(r,\omega) =
Z^{H}_{lm}(r,\omega)R^{\infty}_{lm}(r,\omega)
 + Z^{\infty}_{lm}(r,\omega)R^{H}_{lm}(r,\omega),
\end{align}
where the functions $Z$ are radial integrals over the source
term:
\begin{subequations}\label{Z}
\begin{align} 
Z^\text{H}_{lm}(r,\omega) &= -\frac{1}{\mathcal{A}_{lm}(\omega)} \int_{r_+}^{r} dr'~
\frac{R^\text{H}_{lm}(r',\omega)}{\Delta'^2} \EuScript{T}_{lm}(r',\omega),& \label{ZH} \\ 
Z^\infty_{lm}(r,\omega) &= -\frac{1}{\mathcal{A}_{lm}(\omega)} \int_{r}^\infty dr'~
\frac{R^\infty_{lm}(r',\omega)}{\Delta'^2} \EuScript{T}_{lm}(r',\omega)\;.& \label{ZI} 
\end{align}
\end{subequations}
The function $\mathcal{A}_{lm}(\omega)$ is given by\footnote{The value 
of $\mathcal{A}$ reported in Ref.~\cite{circular} has the wrong sign.}
\begin{equation}
\mathcal{A}_{lm}(\omega) = 2i\omega B^\text{in}_{lm}(\omega)D^\infty_{lm}(\omega).
\end{equation}
This construction of $Z^{\text{H},\infty}$ and ${\cal A}$ follows from
the theory of Green's functions {\cite{Arfken}}.

In this paper, we only evaluate the perturbed field $\psi_4$ [and
therefore the functions $Z$ (\ref{Z})] outside the orbital torus ---
$r < r_{\min}$ and $r > r_{\max}$.  Evaluating $\psi_4$ at arbitrary
locations within the orbital torus $r_{\min} < r < r_{\max}$ would
require a more complicated numerical apparatus than the one developed
here.

\subsection{Bound test mass sources}
\label{ss:Bound test mass sources}

Beginning from Teukolsky's original expression for the source term
$\EuScript{T}$ \cite{Teukolsky}, Breuer \cite{Breuer} has computed the
explicit form of the projected source (see Ref.~\cite{Sasaki Tagoshi}
or Sec.~IV C of Ref.~\cite{circular}):
\begin{equation} \label{source}
\EuScript{T}_{lm}(r,\omega) = \int d\Omega~dt~\EuScript{B}_m(t,r,\theta,\phi,\omega)S_{lm}(\theta,\omega)
e^{i\omega t - im\phi},
\end{equation}
where
\begin{eqnarray}
\EuScript{B}_m &=&
  \sqrt{2} \Delta^2 \rho^3 L_{-1}[\rho^{-4}\bar\rho^2J_+(\rho^{-2}\bar\rho^{-2}\Delta^{-1}T_{n\bar m})] \nonumber \\
&&-2 \rho^3 L_{-1}[\rho^{-4}L_0(\rho^{-2}\bar\rho^{-1}T_{nn})] \nonumber \\
&&+\sqrt{2} \Delta^2 \rho^3 J_+[\rho^{-4}\bar\rho^2\Delta^{-1}L_{-1}(\rho^{-2}\bar\rho^{-2}T_{n\bar m})] \nonumber \\
&&-\Delta^2 \rho^3 J_+[\rho^{-4}J_+(\rho^{-2}\bar\rho T_{\bar m\bar m})]\;.
\end{eqnarray}
This function is given in terms of the tetrad components of the
orbiting particle's energy-momentum tensor,
\begin{equation}
T_{ab}= T^{\alpha\beta}a_\alpha b_\beta\;,
\end{equation}
where $T^{\alpha\beta}$ is the energy-momentum tensor of the particle,
and $a$ and $b$ are either $n$ or $\bar m$.  We define the derivative
operators
\begin{eqnarray}
J_+&=& \partial_r + i K/\Delta\\
L_s&=& \partial_\theta + m\csc\theta - a\omega \sin\theta + s\cot\theta.
\end{eqnarray}
In Kerr spacetime, a point particle of mass $\mu$ has an
energy-momentum tensor given by
\begin{equation} \label{energy momentum}
T^{\alpha\beta}(t,r,\theta,\phi) = \frac{\mu u^\alpha u^\beta}{\Sigma \sin\theta} \frac{d\tau}{dt}
\delta[r-r(t)]\delta[\theta-\theta(t)]\delta[\phi-\phi(t)],
\end{equation}
where $u^\alpha$ is the particle's 4-velocity.  The quantities $r(t)$,
$\theta(t)$, and $\phi(t)$ appearing in Eq.~(\ref{energy momentum})
are the coordinates of the geodesic at coordinate time $t$, and should
not be confused with $r$, $\theta$, and $\phi$ in the definition of
the source term (\ref{source}).

We use the Newman-Penrose basis given by Kinnersly
\cite{Kinnersly} [see Eqs.~(\ref{K vectors}) and (\ref{K 1forms})].
This means the tetrad components of $T^{\alpha\beta}$ are given by
\begin{equation} \label{tradition}
T_{ab} = \frac{C_{ab}}{\sin\theta}\delta[r-r(t)]\delta[\theta-\theta(t)]\delta[\phi-\phi(t)],
\end{equation}
where\footnote{In Ref.~\cite{Minoetalreview}, the $d\theta/d\tau$
terms are missing from the expressions for $C_{n\bar m}$ and $C_{\bar
m \bar m}$.  Also, in Ref.~\cite{equatorial} $d\theta/d\lambda$ is
erroneously replaced with $(d\theta/d\lambda)^2$.} \cite{Sasaki
Tagoshi}
\begin{subequations} \label{C's}
\begin{align}
C_{nn}&= \frac{d\lambda}{dt}\frac{\mu}{4\Sigma^2}
    \left[ E\left(r^2+a^2\right) - aL_z + \frac{dr}{d\lambda} \right]^2,&  
\label{Cnn}\\
C_{\bar m\bar m}&= \frac{d\lambda}{dt}\frac{\mu\rho^2}{2}
    \left[i\left(aE-\frac{L_z}{\sin^2\theta}\right)\sin\theta + \frac{d\theta}{d\lambda}\right]^2,& 
\label{Cnm}\\
C_{n \bar m}&= \frac{d\lambda}{dt} \frac{\mu\rho}{2\sqrt{2}\Sigma}
    \left[ E\left(r^2+a^2\right) - aL_z + \frac{dr}{d\lambda}\right]&
\nonumber \\
& \times  \left[ i\left(aE-\frac{L_z}{\sin^2\theta}\right)\sin\theta + \frac{d\theta}{d\lambda}\right].& 
\label{Cmm}
\end{align}
\end{subequations}
These functions (\ref{C's}) depend explicitly on $dr/d\lambda$ and
$d\theta/d\lambda$, rather than on their squares $V_{r,\theta}$.
Extra bookkeeping is thus needed to keep track of the signs of
$dr/d\lambda$ and $d\theta/d\lambda$ when evaluating them numerically.

Following the details shown in (for example) Ref.~\cite{circular}, we
can now write the projected source term (\ref{source}) in the
following form
\begin{eqnarray} \label{TandA}
\EuScript{T}_{lm}(r,\omega)&=&\int dt~\Delta^2
    \{[A_{nn0} + A_{n\bar m 0} + A_{\bar m \bar m 0}] \delta[r-r(t)]  \nonumber\\
&&+ \partial_r[ (A_{n\bar m 1} + A_{\bar m \bar m 1}) \delta(r-r\{t\})] \nonumber\\
&&+ \partial^2_r[A_{\bar m \bar m 2}\delta(r-r\{t\})] \}e^{i\omega t - im\phi(t)}\;.
\end{eqnarray}
The quantities $A_{abc}$ are shown explicitly in Appendix
\ref{s:Perturbation details}.

Substituting Eq.~(\ref{TandA}) into Eq.~(\ref{Z}) and eliminating the
radial delta functions gives
\begin{eqnarray} \label{ZA}
Z_{lm}^\star(r,\omega) &=& -\frac{1}{A} \int dt~ e^{i\omega t -
im\phi(t)} \Theta^{\star}[r,r(t)]\nonumber\\ && \left[
(A_{nn0} + A_{n\bar m 0} + A_{\bar m \bar m 0}) \right. \nonumber\\ &&
- (A_{n\bar m 1} + A_{\bar m \bar m 1}) \frac{d}{dr} + \left. A_{\bar
m \bar m 2} \frac{d^2}{dr^2}\right] R_{lm}^\star, \nonumber\\
\end{eqnarray}
where $\star = \text{H},\infty$.  We define the step functions
\begin{equation}
\Theta^\infty(x_1,x_2) = \Theta^\text{H}(x_2,x_1) = \Theta(x_2 -x_1),
\end{equation}
in terms of the Heaviside step function $\Theta(x)$.  All other
functions of $r$ and $\theta$ under the integral in Eq.~(\ref{ZA}) are
to be evaluated at $r(t)$ and $\theta(t)$ respectively.  Because $r$
only appears inside the step functions, the quantities
$Z^{\text{H},\infty}$ are independent of $r$ for all $r > r_{\max}$
and $r < r_{\min}$.

To clean up this expression, we now absorb most of the integrand into
a single function:
\begin{align} \label{short}
Z^\star_{lm}(r,\omega) =\int dt~e^{i\omega t - im\phi(t)}
I_{lm}^\star(t,r,\omega)\;.
\end{align}
Following the arguments in Sec.\ V B of Ref.\ {\cite{Drasco Hughes}},
we exploit the harmonic structure of the geodesics (Sec.~\ref{s:Bound
black hole orbits}) to simplify Eq.~(\ref{short}).  In the end, we
will arrive at a general expression for $\psi_4$ (\ref{psi_4}) as a
discrete sum over frequencies, rather than an integral. First we
insert
\begin{align}
&t = \Gamma \lambda + \Delta t,&
&\phi = \Upsilon_\phi \lambda + \Delta \phi,&
\end{align}
into Eq.~(\ref{short}), and we change the integration variable from
$t$ to $\lambda$.  The result is
\begin{eqnarray} \label{new short}
Z^\star_{lm}(r,\omega) = \int d\lambda~ e^{i(\omega\Gamma - m\Upsilon_\phi)\lambda}
J_{lm}^\star(\lambda,r,\omega) ,
\end{eqnarray}
where 
\begin{align}  \label{J}
J_{lm}^\star(\lambda,r,\omega) = \frac{dt}{d\lambda} 
e^{i\omega \Delta t - im\Delta\phi}
I_{lm}^\star(\lambda,r,\omega)\;.
\end{align}
The function $J_{lm}^\star(\lambda,r,\omega)$ is biperiodic, so we can
write it as \cite{Drasco Hughes}
\begin{equation} \label{J exp}
J_{lm}^\star(\lambda,r,\omega) = \sum_{kn} J^{\star}_{lmkn} e^{-i(k \Upsilon_\theta + n \Upsilon_r)\lambda}\;;
\end{equation}
the constants $J^\star_{lmkn}$ are given by 
\begin{eqnarray}
J^\star_{lmkn} &=& \frac{1}{(2\pi)^2}\int_0^{2\pi}dw_\theta \int_0^{2\pi}dw_r~e^{i(k w_\theta + n w_r)} 
\nonumber \\ 
&&\times J^\star_{lm}[r(w_r),\theta(w_\theta),r,\omega] . 
\end{eqnarray}
By inserting the expansion (\ref{J exp}) into Eq.~(\ref{new short}),
we find that the integral (\ref{new short}) is just a sum of delta
functions:
\begin{equation}  \label{zdelta}
Z^\star_{lm}(r,\omega) = \sum_{kn} Z^\star_{lmkn}(r) \delta(\omega -
\omega_{mkn})\;.
\end{equation}
We have used the coordinate time frequencies (\ref{t frequencies}) to
define
\begin{equation}
\omega_{mkn} = m\Omega_\phi + k\Omega_\theta + n\Omega_r. 
\end{equation}
The expansion coefficients for $Z^\star_{lm}(r,\omega)$ are given by
\begin{align} \label{long Zlmnk}
Z^\star_{lmkn}(r) = 
&\frac{1}{2\pi\Gamma}\int_0^{2\pi}dw_\theta~\int_0^{2\pi} dw_r~e^{ikw_\theta + inw_r}&
\nonumber\\ 
&\times J^\star_{lm}(w_r,w_\theta,r,\omega_{mkn}),&
\end{align}
By substituting the expanded form (\ref{zdelta}) of $Z$ into the
general expression (\ref{psi_4}) for $\psi_4$, we eliminate the
frequency integral to obtain
\begin{equation}
\psi_4 = \rho^4 \sum_{lmkn} R_{lmkn}(r)S_{lmkn}(\theta)
e^{-i\omega_{mkn}t+im\phi} \label{psi exp}\;.
\end{equation}
The radial and angular functions are now discrete functions of
frequency, inheriting the angular harmonic index $k$ and the radial
harmonic index $n$:
\begin{align}
S_{lmkn}(\theta) = &S_{lm}(\theta,\omega_{mkn}),& \\
R^\star_{lmkn}(r) = &R^\star_{lm}(r,\omega_{mkn}),&  \\
R_{lmkn}(r) = &Z^\text{H}_{lmkn}(r)R^{\infty}_{lmkn}(r)
                   + Z^{\infty}_{lmkn}(r)R^\text{H}_{lmkn}(r).
\label{eq:greens}& 
\end{align}

\subsection{Numerical considerations}
\label{ss:Numerical considerations}

Because the functions $r(w_r)$ and $\theta(w_\theta)$ are periodic and
even [cf.\ Eq.\ (\ref{even coords})], the range of the integrals
(\ref{long Zlmnk}) can be reduced by a factor of two.  Writing the
integrand in (\ref{long Zlmnk}) as $z(w_r,w_\theta)$, we can expand
the integrals into the form
\begin{align}
Z = 
&\left[ \int_0^{\pi}\!dw_\theta \int_0^{\pi} \!dw_r 
 + \int_0^{\pi}\!dw_\theta \int_\pi^{2\pi} \!dw_r \right.& 
\nonumber \\
&\left.\!\!+ \int_\pi^{2\pi}\!dw_\theta \int_0^{\pi} \!dw_r 
 + \int_\pi^{2\pi}\!dw_\theta \int_\pi^{2\pi} \!dw_r \right] z(w_r,w_\theta).&
\end{align}
Since the integrand has a period of $2\pi$ in both variables, we can
shift all of the $\{\pi,2\pi\}$ branches to $\{-\pi,0\}$:
\begin{align}
Z = 
&\left[ \int_0^{\pi}dw_\theta \int_0^{\pi} dw_r 
 + \int_0^{\pi}dw_\theta \int_{-\pi}^{0} dw_r \right.& 
\nonumber \\
&\left.\!\!+ \int_{-\pi}^0dw_\theta \int_0^{\pi} dw_r 
 + \int_{-\pi}^0 dw_\theta \int_{-\pi}^0 dw_r \right] z(w_r,w_\theta).&
\end{align}
We can now reflect all of the $\{-\pi,0\}$ branches across the origin to get
\begin{align} \label{short z}
Z = \int_0^{\pi}dw_\theta \int_0^{\pi} dw_r \sum_{D_r = \pm} \sum_{D_\theta = \pm}
z(D_r w_r,D_\theta w_\theta)\;.
\end{align}

Direct evaluation of Eq.~(\ref{short z}) is complicated by the fact
that it is difficult to evaluate $r(w_r)$ and $\theta(w_\theta)$ [see
discussion preceeding Eq.~(\ref{final delta})].  We avoid this problem
by changing integration variables from $w_r$ and $w_\theta$ to $\psi$
and $\chi$ [see Eq.~(\ref{new coords})].

Eccentric orbits introduce another numerical nuance. The integrand in
Eq.~(\ref{short z}) is moderately divergent if the orbit covers a
large range in $r$ (i.e., if the orbit is highly eccentric).  This
problem can be traced to the scaling behavior (\ref{inup}) of
solutions to the homogeneous radial equation.  In particular, the
typical size of the integrand at $r_{\min}$ can be different from
typical values at $r_{\max}$ by orders of magnitude.  This causes an
artificial focusing of numerical accuracy.  Our choice of boundary
conditions (\ref{initial rt}) is such that the turning points
$r_{\min}$ and $r_{\max}$ occur at $\psi = 0$ and $\psi = \pi$
respectively.  To avoid this problem, we introduce a change of
integration variables from $\psi$ to $\zeta$ where $d\psi/d\zeta =
e^{-\psi/\Psi^\star}$, with $\star = \text{H},\infty$, and
\begin{align}
\zeta(\psi) = \Psi^\star \left( e^{\psi/\Psi^\star} - 1\right).
\end{align}
Here $\Psi^\star$ is a constant (which can be positive or negative).
Before integrating, we tune the value of $\Psi^\star$ so that the
typical magnitude of the integrand at the outer turning point is
comparable to that at the inner turning point.  In general, the value
of $\Psi^\star$ will decrease if either $l$ or eccentricity $e$ are
increased.

Applying these various changes of variable and massaging of the
integration ranges to Eq.~(\ref{long Zlmnk}), our final expression for
the $Z^\star_{lmkn}(r)$ coefficients is given by
\begin{align} \label{final Z}
Z^\star_{lmkn}(r) = 
&\frac{1}{2\pi\Gamma}\int_0^{\pi}d\chi \int_0^{\zeta(\pi)}\!\! d\zeta
\sum_{D_r = \pm} \sum_{D_\theta = \pm}\!\!
\frac{dw_\theta}{d\chi}\frac{dw_r}{d\psi}\frac{dt}{d\lambda}&
\nonumber \\
&\times  e^{-\psi /\Psi^\star} I^\star_{lm}\left(w_r, w_\theta, D_r, D_\theta, r, \omega_{mkn} \right)&
\nonumber \\
&\times \exp \left[iD_\theta \left( kw_\theta + \omega_{mkn} \Delta t^\theta - m \Delta \phi^\theta \right)\right]&
\nonumber \\
&\times \exp \left[iD_r \left( nw_r + \omega_{mkn} \Delta t^r - m \Delta \phi^r \right)\right],&
\end{align}
where $I^\star_{lm}$ is defined by Eq.~(\ref{short}).  The functions
$r(t)$, $\theta(t)$ and their derivatives are evaluated as
\begin{align}
&r(t) \to r(\zeta),&
&\frac{dr}{d\lambda}[r(t)] \to D_r\left| \frac{dr}{d\lambda}[r(\zeta)] \right|,&
\nonumber \\
&\theta(t) \to \theta(\chi),& 
&\frac{d\theta}{d\lambda}[\theta(t)] \to D_\theta\left| \frac{d\theta}{d\lambda}[\theta(\chi)] \right|.&
\end{align}

Lastly, note that applying the simultaneous transformations $m\to -m$,
$\omega\to -\omega$ to the radial (\ref{Teukolsky}) and the angular
equations (\ref{angular}), one can derive a simple symmetry rule for
expansion coefficients \cite{circular,equatorial}
\begin{equation} \label{Zsym}
Z^\star_{l(-m)(-k)(-n)}(r) = (-1)^{l+k} \bar Z^\star_{lmkn}(r).
\end{equation}
This symmetry is used to reduce the computation time for various
summations (see Sec.~\ref{ss:Truncation}).

\subsection{Waveforms and fluxes}
\label{ss:Waveforms and fluxes}

In this section, we use the expanded form of the curvature
perturbation (\ref{psi exp}) to derive expressions for the leading
order metric perturbation at infinity, and for the radiated fluxes of
energy and angular momentum to infinity and down the black hole's
event horizon.  Since we will be interested in quantities at both
infinity and the horizon, it will be useful to first define the
limiting values of the coefficients $Z_{lmkn}^\star(r)$ as follows:
\begin{subequations}\label{lim Zs}
\begin{align} 
&Z_{lmkn}^\text{H} = D^\infty_{lmkn}  Z_{lmkn}^\text{H}(r > r_{\max}), &
\label{lim ZH}
\\
&Z_{lmkn}^\infty = B^\text{hole}_{lmkn}  Z_{lmkn}^\infty(r < r_{\min}).&
\label{lim ZI}
\end{align}
\end{subequations}
Recall that the functions $Z_{lmkn}^\star(r)$ are independent of $r$ for all
$r > r_{\max}$ and $r < r_{\min}$.

At infinity, the leading order curvature and metric perturbations are
related by
\begin{equation} \label{curvature perturbation}
\psi_4(r \to \infty) = \frac{1}{2}\frac{\partial^2}{\partial t^2}
\left( h_+ - ih_\times \right).
\end{equation}
Here $h_+$ and $h_\times$ are the two independent components of the metric 
perturbation, defined by 
\begin{equation} \label{metric perturbation}
h_{\alpha\beta}(r\to\infty) = h_+ e^+_{\alpha\beta} + h_\times e^\times_{\alpha\beta} + O(1/r^2)\;,
\end{equation}
where $e^{+}_{\alpha\beta}$ and $e^{\times}_{\alpha\beta}$ are
polarization tensors \cite{Blandford Thorne}.  If we now substitute
the expanded form of the curvature perturbation (\ref{psi exp}) into
the left hand side of Eq.~(\ref{curvature perturbation}), evaluate the
limit, and integrate twice in coordinate time (setting the arbitrary
integration constants to zero), we find
\begin{align} \label{waveform}
h_+ - i h_\times = -\frac{2}{r}\sum_{lmkn} \frac{Z_{lmkn}^\text{H}}{\omega_{mkn}^{2}}
                   S_{lmkn}(\theta) e^{-i\omega_{mkn}(t-r)+im\phi}\;.
\end{align}
We have used the definition (\ref{lim ZH}) for the limiting value of
$Z_{lmkn}^\text{H}(r)$.

It is important to note that the expression for $Z^\text{H}_{lmkn}$
derived above [e.g. Eq.~(\ref{final Z})] assumed geodesics with a
specific choice of initial position: $r(0) = r_{\min}$, $\theta(0) =
\theta_{\min}$, $t(0) = 0$, and $\phi(0) = 0$.  These are the fiducial
geodesics from Ref.~\cite{Drasco Flanagan Hughes}.  Making a different
choice of initial position will result in an overall phase change for
$Z^\text{H}_{lmkn}$.  The details of this phase can be found in
Ref.~\cite{Drasco Flanagan Hughes}.  As we will see below, the
expression for the waveform (\ref{waveform}) is unique in that, unlike
the formulas for evolving $E$, $L_z$, it depends explicitly on this
overall phase, and therefore on the initial position of the geodesic.

At infinity, the effective energy-momentum tensor
$T^\text{GW}_{\alpha\beta}$ is easily built from the GWs
\cite{Isaacson}. The result can be expressed as \cite{Blandford
Thorne}
\begin{equation}
T^\text{GW}_{\alpha\beta} = \frac{1}{16\pi}
\left< 
\frac{\partial h_+}{\partial x^\alpha} \frac{\partial h_+}{\partial x^\beta}
+ 
\frac{\partial h_\times}{\partial x^\alpha} \frac{\partial h_\times}{\partial x^\beta}
\right>\;.
\end{equation}
Brackets indicate an average over several wavelengths of the
gravitational waves.  Using this result and the expansion
(\ref{waveform}), it is straightforward to show that waves carry an
averaged flux of energy and angular momentum given by
\begin{subequations}
\begin{align}
&\left< \frac{dE}{dt} \right>^\infty = \sum_{lmkn} \frac{1}{4\pi\omega_{mkn}^2} \left| Z_{lmkn}^\text{H}\right|^2,& \\
&\left< \frac{dL_z}{dt} \right>^\infty = \sum_{lmkn}\frac{m}{4\pi\omega_{mkn}^3} \left| Z_{lmkn}^\text{H}\right|^2.& 
\end{align}
\end{subequations}
Similar expressions can be found for the flux of energy and angular
momentum through the horizon; we refer the reader to
Refs.~\cite{Poisson PDE,teukpress,circular} for a detailed derivation.
The resulting fluxes are
\begin{subequations} \label{Horizon fluxes}
\begin{align}
&\left< \frac{dE}{dt} \right>^\text{H} = \sum_{lmkn} \frac{1}{4\pi\omega_{mkn}^2} \alpha_{lmkn}\left| Z_{lmkn}^\infty\right|^2,& \\
&\left< \frac{dL_z}{dt} \right>^\text{H} = \sum_{lmkn}\frac{m}{4\pi\omega_{mkn}^3}\alpha_{lmkn} \left| Z_{lmkn}^\infty\right|^2.& 
\end{align}
\end{subequations}
The superscripts we use here are somewhat confusing: the fluxes to
infinity depend upon the {\it horizon} coefficients,
$Z^\text{H}_{lmkn}$, and the fluxes down the horizon depend upon the
{\it infinity} coefficients $Z^\infty_{lmkn}$.  This seemingly obtuse
convention comes from the Green's function solution we constructed,
Eq.\ (\ref{eq:greens}).  The coefficients $Z^\text{H}$ set the
amplitude of the radial behavior towards infinity; the coefficients
$Z^\infty$ set the amplitude towards the horizon.

The coefficient $\alpha_{lmkn}$ appearing in Eq.\ (\ref{Horizon
fluxes}) is given by \cite{circular}
\begin{equation}
\alpha_{lmkn} = \frac{256 (2Mr_+)^5 P(P^2 + 4\epsilon^2)(P^2 +
16\epsilon^2)\omega_{mkn}^3}{|C_{lmkn}|^2}\;,
\label{eq:horizonalphadef}
\end{equation}
with $\epsilon = \sqrt{M^2 - a^2}/4Mr_+$, where $r_+$ is the location of the event horizon, and
\begin{align}
|C_{lmkn}|^2 &= \left[(\lambda_{lmkn}+2)^2 + 4 a \omega_{mkn} - 4 a^2 \omega_{mkn}^2\right]& \nonumber \\
             &\times\left(\lambda_{lmkn}^2 + 36 m a \omega_{mkn} - 36 a^2\omega_{mkn}^2\right)& \nonumber\\
             &+ \left(2\lambda_{lmkn}+3\right) \left(96 a^2\omega_{mkn}^2 - 48 m a \omega_{mkn}\right) & \nonumber\\
             &+ 144 \omega_{mkn}^2(M^2 - a^2).&
\end{align}
Recall that $P = \omega_{mkn} - ma/(2Mr_+)$, and that $\lambda_{lmkn}$
is related to angular equation's eigenvalue [cf.\ Eq.\
(\ref{angular})] via Eq.~(\ref{lambda eig}) with $\omega =
\omega_{mkn}$.

To go from snapshot waveforms to adiabatic inspiral waveforms, we must
evolve not just energy and angular momentum, but also the Carter
constant $Q$.  This will eventually be done with the formalism
recently derived by Mino \cite{Mino}.  However, if we assume that
radiation does not influence an orbit's inclination [$\iota$ or
equivalently $\theta_\text{inc}$], we can can infer an expression for
$\left< dQ/dt \right>$.  This scheme\ was first suggested by Curt
Cutler \cite{Cutler private}.  From the definition (\ref{iota}) of
$\iota$, we see that setting $\left< d\iota/dt \right> = 0$ gives
\begin{equation}\label{Qdot}
\left< \frac{dQ}{dt} \right> = \frac{2Q}{L_z} \left< \frac{dL_z}{dt} \right>,
\end{equation}
where $\left< dL_z/dt \right> = \left< dL_z/dt \right>^\text{H} +
\left< dL_z/dt \right>^\infty$.  Hughes showed that this approximation
is reasonable in the case of circular orbits \cite{circularII}.  It is
currently unknown how accurate it may be in the case of generic orbits
(though it is almost certainly {\it pathological} for orbits with
nearly vanishing $L_z$ {\cite{gair_private}}).  In
Sec.~\ref{s:Results} below, we report values of $\left< dQ/dt \right>$
given by Eq.~(\ref{Qdot}).  We should emphasize however that, since
Eq.~(\ref{Qdot}) is itself a speculation, those results will very
likely be less accurate than the associated fluxes of energy and
angular momentum.

\section{Numerical algorithm}
\label{s:Numerical algorithm}

For a given orbit, the calculation of the snapshot waveform is
essentially identical to the calculation of each of the radiative
fluxes.  Given an engine which can compute the $Z_{lmkn}^\star$
coefficients, which we will refer to simply as modes, each calculation
is essentially just a long four dimensional summation.  In the
following three subsections, we describe the details of our algorithms
for (i) computing the frequency domain representations of the
geodesics, (ii) computing the individual modes $Z_{lmkn}^\star$, and
(iii) truncating the sums representing the waveforms and fluxes.

\subsection{Geodesics}
\label{ss:Geodesics}

We begin by computing the quantities associated with the specified
geodesic for a given set of black hole and orbital parameters
$(a,e,p,\theta_\text{inc})$.  This calculation proceeds as follows:
\begin{enumerate}

\item Compute the constants of the motion $E$, $L_z$, $Q$

\item Compute the orbital frequencies $\Upsilon_{r,\theta,\phi}$, and
  $\Gamma$

\item Construct routines to evaluate $w_r(\psi)$, $w_\theta(\chi)$ and
  their derivatives

\item Construct the expansions (\ref{expanded x}) of $\Delta t$ and
  $\Delta \phi$

\end{enumerate}
Explicit expressions for the first three of these steps are shown in
Appendix \ref{s:Geodesic details}.  The last step requires that we
approximate the sums for the expansions (\ref{expanded x}) of $\Delta
x= \Delta t,\Delta \phi$ as
\begin{align}
&\Delta x^\theta \approx \sum_{k=1}^K \Delta x_k^\theta \sin(k w_\theta),&
&\Delta x^r \approx \sum_{n=1}^N \Delta x_n^r \sin(n w_r),&
\end{align}
where we determine the values of $K$ and $N$ by requiring that 15
consecutive $\Delta x_{k,n}^{r,\theta}$ coefficients make fractional
contributions to the sum which are less than some specified accuracy;
we typically find $K,N\lesssim 50$.  Each of these calculations
associated with the geodesics is done to a fractional accuracy of
$10^{-12}$, and the results are made available to the remaining
computations.  The calculations described so far are typically
completed in a few seconds (using a single $\sim 1$ GHz processor).
It's worth bearing in mind that these calculations need only be done
once when computing the waveform and fluxes for a given orbit.

\subsection{Modes}
\label{ss:Modes}

After computing geodesic quantities, we construct an engine for
calculating the modes $Z^\star_{lmkn}$.  This requires solutions to
the angular equation, and to the homogeneous radial equation.  (Our
homogenenous solutions have an implicit coupling to the source, since
they depend on the frequencies $\omega_{mkn}$, which are determined by
the orbit.)  The routines for solving the homogeneous radial and
angular equations were inherited from the code used in
Ref.~\cite{circular}, where they are described in detail.  We solve
the angular equation to nearly machine accuracy, and we solve the
radial equation to a fractional accuracy of $\min(10^{-7},
\varepsilon_\text{flux}/100)$, where $\varepsilon_\text{flux}$ is the
overall fractional accuracy demanded of the waveforms and fluxes.

When computing (\ref{final Z}), we treat the radial integral as the
outermost integral [the reverse of how we have written Eq.~(\ref{final
Z})].  After experimenting with a variety of numerical methods for
evaluating these integrals, we found a Clenshaw-Curtis quadrature
algorithm to be the most efficient.  This method analytically
integrates a numerical series representation (in Chebyshev
polynomials) of the integrand, to produce a series representation of
the integral (for more details, see for example Ref.~\cite{NR}). Each
of these calculations depends on the indices $(l,m,k,n)$, and must be
repeated many times to compute the waveform and fluxes associated with
a given orbit.

We compute the angular integral in Eq.~(\ref{final Z}) to a fractional
accuracy of $\min(10^{-6},\varepsilon_\text{flux}/10)$.  The accuracy
demanded of the radial integral is chosen dynamically.  We begin by
asking for a fractional accuracy of
$\min(10^{-5},\varepsilon_\text{flux})$.  As each mode is computed, we
store the magnitude of the largest modes encountered so far.  Later
modes are then computed to a fractional accuracy of at least 10\%.
After this 10\% accuracy has been achieved the integrator continues to
add more terms to the Chebyshev expansion.  At each iteration it
estimates the associated uncertainty in the total energy fluxes (the
energy flux at the horizon for the $Z^\infty_{lmkn}$ integrals, and
the energy flux at infinity for the $Z^\text{H}_{lmkn}$ integrals) by
comparing to the magnitudes of the largest known modes.  It then
truncates the Chebyshev expansion as soon as the integral's associated
relative flux error is smaller than $\varepsilon_\text{flux}$.  The
reason for this dynamic accuracy control is that many of the modes are
very near zero.  If we know that the total flux contains modes which
are $\sim 10^{-3}$, there is no need to compute a mode which is $\sim
10^{-20}$ beyond the first couple of digits.  A single $\sim 1$ GHz
processor usually takes considerably less than a second to compute
both $Z^\infty_{lmkn}$ and $Z^\text{H}_{lmkn}$ (see Fig.~1 in
Ref.~\cite{Hughes et al}), unless the modes are especially badly
behaved in the sense described in Sec.~\ref{ss:Truncation} below.  In
these rare cases the modes can take anywhere from 10 seconds to a
minute to compute.

\begin{figure}
\includegraphics[width = .45\textwidth]{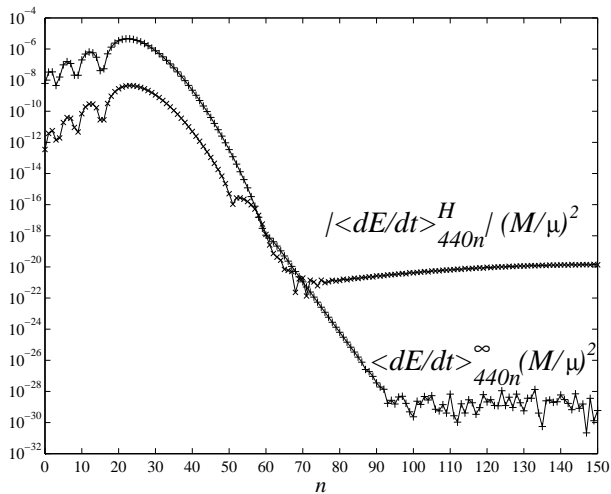}
\caption{Modal energy flux through the horizon (superscript H, marked
with $\times$'s) and at radial infinity (superscript $\infty$, marked
with $+$'s).  Results are for $l = m = 4$, $k = 0$.  The orbit has
eccentricity $e = 0.7$, inclination $\theta_\text{inc} = 45^\circ$,
and semilatus rectum $p = 5.1$, while the massive black hole has $a =
0.9M$.}
\label{f:diverge}
\end{figure}

When computing a flux, the majority of the numerical work is devoted
to computing modes which are insignificant or zero.  The integrator
has little difficulty when computing the dominant modes.  However,
modes which are very near zero (near round off error) can cause the
integrator to fail.  This typically happens when computing modes with
``large'' values of the indices ($l,m,k,n$); the meaning of ``large''
is somewhat orbit and index dependent.  For example, for fixed
($l,m,k$), we find that there is some limiting value of $n$ beyond
which the mode calculations cannot be trusted.

We show an example of this behavior in Fig.~\ref{f:diverge}.  Beyond
the dominant modes at $n \simeq 20$, the mode magnitudes fall
(roughly) exponentially until reaching a bottom.  For the case shown
in Fig.~\ref{f:diverge}, the horizon modes bottom out at about $n=70$,
after which they gradually diverge, while the infinity modes bottom
out at about $n=90$, after which they appear to be dominated by
numerical error.  Both behaviors are unphysical since they indicate a
divergent total energy flux.  The different behaviors arise from the
different scalings of these modes with frequency:
\begin{eqnarray}
\left\langle\frac{dE}{dt}\right\rangle^\infty_{lmkn} \sim
\omega_{mkn}^{-2} |Z^\text{H}_{lmkn}|^2\;,
\label{Einf scaling}\\
\left\langle \frac{dE}{dt}\right\rangle^\text{H}_{lmkn} \sim
\omega_{mkn}^2 |Z^\infty_{lmkn}|^2\;.
\label{EH scaling}
\end{eqnarray}
This difference comes from the factor $\alpha_{lmkn}\sim
\omega_{mkn}^4$ in Eq.~(\ref{Horizon fluxes}).

For reasonable requested accuracies, the modes which are difficult to
compute are always insignificant when compared to the dominant modes
(as is clearly the case in Fig.~\ref{f:diverge}).  In the vast
majority of cases, the truncation rules (see Sec.~\ref{ss:Truncation}
below) have been invoked long before these badly behaved modes are
encountered.  If this is not the case, these badly behaved modes can
undermine the truncation scheme since, despite their small magnitudes,
they do not regularly decay with increasing index.  In order to
account for this, we artificially zero any modes for which the ratio
of the magnitude of the integral $Z^\star_{lmkn}$ to the maximum over
$\zeta$ and $\chi$ of the absolute value of the integrand in
Eq.~(\ref{final Z}) is below some threshold.  We have found that this
ratio is a good indicator of when the modes have bottomed out in the
sense shown in Fig.~\ref{f:diverge}.  Unfortunately, we have had to
``hand-tune'' this parameter somewhat. This was particularly so as we
explored orbits with increasing eccentricity.  In the future, we
expect to incorporate our code into one which computes inspiral
waveforms.  Hand tuning of this parameter will be unacceptable for
such a code, and a more robust technique will be needed.

\subsection{Truncation}
\label{ss:Truncation}

Given the routines described above, the remaining task is to evaluate
the four dimensional sums which make up the waveforms and fluxes
(Sec.~\ref{ss:Waveforms and fluxes}).  As mentioned above, we do this
by monitoring the sums which represent the fluxes.  We assume that
once the fluxes have converged to some requested fractional accuracy
$\varepsilon_\text{flux}$, the sums representing the waveforms are
comparably accurate since they are constructed from the same mode
coefficients as were used for the fluxes.  We now discuss the
algorithm for truncating the sums which represent the fluxes.

Let $F$ be the flux of energy or angular momentum at the horizon or at
infinity
\begin{equation}
F = \left< \frac{dE}{dt} \right>^{\text{H},\infty}
  , \quad \left< \frac{dL_z}{dt} \right>^{\text{H},\infty}.
\end{equation}
The algorithm for computing these fluxes is:
\begin{subequations}\label{sum outline}
\begin{align}
F &= \sum_{l=2}^\infty F_l,& 
\\
F_l &= \sum_{m=-l}^l F_{lm},& 
\\
F_{lm} &= \sum_{k=-\infty}^\infty F_{lmk},&
\\
F_{lmk} &= F_{lmk0} + 2\sum_{n=1}^\infty F_{lmkn}.&
\end{align}
\end{subequations}
We have used the symmetry (\ref{Zsym}), which gives
\begin{equation}
F_{lmkn} = F_{l(-m)(-k)(-n)},
\end{equation}
to simplify the sum over the radial harmonic index $n$. Each of the
sums which has an infinite boundary is then truncated as described
below.
  
The terms which make up outermost sum, $F_l$ in the sum over $l$,
always decrease monotonically with increasing $l$.  This means that
the $l$-sum can be written
\begin{equation}
F = \sum_{l=2}^L F_l + \delta F,
\end{equation}
where to a good approximation $\delta F \approx F_L$, and $L$ is a
number which is determined by the requested accuracy via
\begin{equation}
|F_L| < \varepsilon_\text{flux} \max_l |F_l|\;.
\end{equation}

The sum over the $m$-index is computed for all values $-l \le m \le l$
--- we do not truncate it.

The $k$ and $n$ sums are approximated as
\begin{align}
F_{lm} &\approx \sum_{k=K_-}^{K_+} F_{lmk},& 
F_{lmk} &\approx F_{lmk0} + 2\sum_{n=1}^{N_+} F_{lmkn},&
\end{align}
where the constants $K_\pm$ and $N_+$ are determined by requiring 
that the following relation
\begin{subequations}
\begin{eqnarray}
|F_{lmK_\pm}| &<& \varepsilon_\text{flux} \max_{l'm'k'} |F_{l'm'k'}|,
\\
|F_{lmkN+}| &<& \varepsilon_\text{flux} \max_{l'm'k'n'} |F_{l'm'k'n'}|,
\end{eqnarray}
\end{subequations}
(where the maximization is done over all previously encountered mode
amplitudes) is satisfied for $B$ consecutive terms.  We also require
that the first of these $B$ terms is larger in magnitude than the
last.  This extra condition gives some assurance that we are not
truncating during a slow rising up of the modes.  For the sums over
the polar harmonic index $k$, we use $B = 2$.  For the sums over the
radial harmonic index $n$, we use $B=5$ --- the dominant radial modes
show a less orderly distribution than the polar modes.

The validity of any given truncation scheme is always subject to doubt
--- it could be the case that modes which haven't been computed are
anomalously large in magnitude.  In this sense truncating sums can be
likened to financial investments: past performance is not necessarily
indicative of future results.

We have found these words of caution are especially applicable for the
sums over the radial harmonic index $n$.  We find that the $F_l$
terms, as defined by Eqs.~(\ref{sum outline}), in the outermost sum
over $l$ always decrease monotonically with increasing $l$.  Also, the
$F_{lmk}$ terms are always peaked near $k=0$, so that setting the
buffer $B$ to a modestly small number is sufficient for catching the
dominant modes.  Unfortunately, the $F_{lmkn}$ appear not to obey any
similar general rules.  If either $l$ or the eccentricity $e$ are
increased, the $n$ values of the dominant $F_{lmkn}$ modes drift to
large values of $n$.  As a result, the truncation rules are expected
to be limited in a rough sense by large eccentricity, or by requests
for high accuracy (which would require modes with large $l$).

\begin{figure}
\includegraphics[width = .45\textwidth]{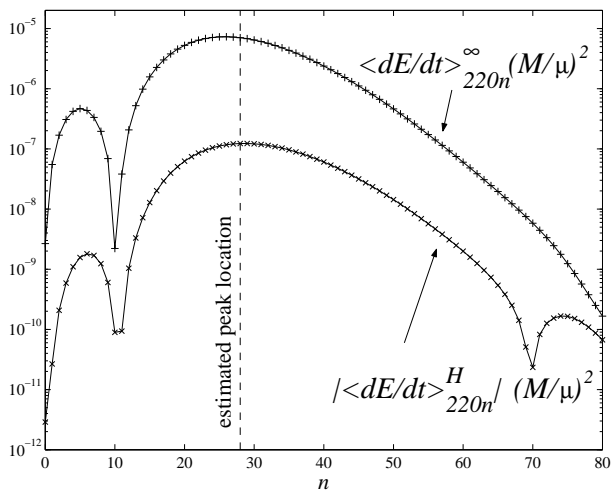}
\caption{Modal energy flux through the horizon (superscript H, marked
with $\times$'s) and at infinity (superscript $\infty$, marked with
$+$'s).  The indices $l = m = 2$, $k=0$.  The orbit has eccentricity
$e = 0.85$, inclination $\theta_\text{inc} = 20^\circ$, and semilatus
rectum $p = 6$, while the massive black hole has $a = 0.9M$.  The
horizon modes are peaked at $n=29$, while the infinity modes are
peaked at $n=26$.  The dashed line shows the estimated peak location
(at $n=28$) found using the Peters-Mathews power formula (\ref{PM
power}).}
\label{f:PM}
\end{figure}

It may be possible to better anticipate $n$ for the dominant modes.
Peters and Mathews \cite{Peters Mathews} used the quadrupole
approximation to describe the radiation of two point particles moving
along closed Keplerian orbits of arbitrary eccentricity $e$.  Under
this approximation, they decomposed the metric perturbation into
tensor spherical harmonics \cite{Thorne} with $l=2$.  They then showed
that the power $\left< dE/dt\right>^{\text{PM}}_n$ radiated at a
frequency $n\omega$ (where $n$ is an integer and $\omega$ is the
single unique orbital frequency for a Keplerian orbit) is given by
\begin{align}\label{PM power}
\left\langle \frac{dE}{dt} \right\rangle^{\text{PM}}_n\!\!
&\propto  \frac{n^4}{32} \biggl\{ \bigl[ J_{n-2}(ne) - 2eJ_{n-1}(ne) &
\nonumber \\
& + \frac{2}{n}J_n(ne)+2eJ_{n+1}(ne)-J_{n+2}(ne) \bigr]^2&
\nonumber \\
&+ (1-e^2)\left[ J_{n-2}(ne) - 2J_n(ne)+J_{n+2}(ne) \right]^2&
\nonumber \\
& + \frac{4}{3n^2}\left[ J_n(ne) \right]^2 \biggr\} ,&
\end{align}
where $J_n(x)$ are Bessel functions of the first kind.  We have found
that this formula (which was derived using only the quadrupole
equation and Keplerian orbits) accurately predicts the $n$ values of
the dominant $F_{lmkn}$ modes for $l=2$.  More quantitatively,
define ${\tilde n}$ by
\begin{equation}
\left< \frac{dE}{dt} \right>^{\text{PM}}_{\tilde n} = \max_n \left<
\frac{dE}{dt} \right>^{\text{PM}}_n ,
\end{equation}
so that
\begin{equation}
\tilde n \approx \exp\left[ \frac{1}{2} -\frac{3}{2} \ln (1-e) \right];
\end{equation}
this approximate formula was found by fitting to numerically
determined maxima of Eq.~(\ref{PM power}).  Then, even for highly
relativistic, inclined, and eccentric orbits, we have
\begin{equation}
\max_n \left| F_{2mkn}\right| \approx \left| F_{2mk\tilde n} \right|.
\end{equation}
An example is shown in Fig.~\ref{f:PM}.
Reproducing Fig.~\ref{f:PM} under the same conditions but with an
orbital inclination of $\theta_\text{inc} = 70^\circ$ shifts the peaks
occur to $n \simeq 35$.  It is somewhat remarkable that the
Peters-Mathews result can be used to predict the peak locations in
this way, since the generic Kerr orbits considered here do not closely
resemble the analogous Keplerian orbits either in their orbital
frequencies or in their shapes.  We note that while the Peters-Mathews
power formula (\ref{PM power}) gives reasonable estimates for the
location in $n$ of the dominant modes, it does not accurately predict
the actual flux associated with the dominant modes (for example, the
top panel of Fig.~14 in Ref.~\cite{equatorial} shows a very narrow
peak when compared to the relatively broad peak for the $e=0.7$ orbit
in Fig.~3 from Ref.~\cite{Peters Mathews}).

It is likely that the $l>2$ analog of the Peters-Mathews power formula
(\ref{PM power}) would be a useful estimator of the values of the
radial harmonic index $n$ for the dominant modes.  Such an estimator
would be valuable tool for designing a more accurate truncation scheme
for the sums over $n$.  The value of $n$ for the dominant modes
$F_{lmkn}$ increases roughly linearly with $l$, so it may be possible
to ``track'' the peak location starting from the Peters-Mathews power
formula (\ref{PM power}) at $l=2$.  Future progress along these lines
may reduce the limitations of our code for computing accurate fluxes
for highly eccentric orbits.

\subsection{Validation}
\label{ss:Validation}

We have compared the output of our code with two existing codes.  The
first, the ``circular code'' by Hughes \cite{circular}, treats orbits
which have a non-zero inclination and a constant radius.  The second,
the ``equatorial code'' by Glampedakis and Kennefick
\cite{equatorial}, treats orbits which are eccentric and confined to
the equatorial plane. Both of these codes have already passed a
variety of tests of their own (described below, but see
Refs.~\cite{circular,equatorial} for more details).  Also, the
circular and equatorial codes have been shown to agree with each other
for orbits which are both circular and equatorial \cite{equatorial}.

The circular code has been shown to agree with analytical
post-Newtonian approximations, found in Ref.~\cite{Minoetalreview},
in the weak field limit.  It has also been shown to have the correct
rotational properties in the Schwarzschild limit (see Sec.~V B of
Ref.~\cite{circular} for details). Our code is a direct descendant of
the circular code, and it has some elements in common with that code.
Namely, both use the same routines for solving the homogeneous radial
equation, and also for solving the angular equation.  However, they
differ in their methods for solving the geodesic equations, in their
methods for computing the quantities $Z^\star_{lmkn}$, and in
their methods for truncating the sums representing the fluxes.  When
computing the individual modes
$\left<dX/dt\right>^{\text{H},\infty}_{lmkn}$ (for $X = E,L_z$) with
our code and the circular code, we find differences $\Delta
\left<dX/dt\right>^{\text{H},\infty}_{lmkn} \lesssim
(10^{-5})\left<dX/dt\right>^{\text{H},\infty} $.  Here we are
comparing the difference to the total flux rather than the mode itself
because often there can be large fractional disagreements for
insignificant modes.  When computing the total fluxes
$\left<dX/dt\right>^{\text{H},\infty}$ with these two codes, we find
differences $\Delta \left<dX/dt\right>^{\text{H},\infty} \lesssim
(10^{-5})\left<dX/dt\right>^{\text{H},\infty}$.

The equatorial code has been tested against its own direct ancestors
\cite{ckp,Kennefick}, against the circular code, and also against
Shibata's earlier equatorial code \cite{Shibata equatorial}.  Note
that in the case of the test against Shibata's code, Glampedakis and
Kennefick found agreement only at a level of about 1\%.  Shibata has
since reported that after improving his code, he now agrees with
Glampedakis and Kennefick to much greater accuracy \cite{Shibata fix}.
When computing the individual modes $\langle
dX/dt\rangle^{\text{H},\infty}_{lmkn}$ (for $X = E,L_z$) our code
always agrees with the equatorial code at the level of the equatorial
code's claimed fractional accuracy of $10^{-3}$. We have tested this
for the top panels of Figs.~14-16 of Ref.~\cite{equatorial}; by
fractional accuracy, we mean the ratio of the differences in modes to
the sum of all the modes shown in each of the figures.  When computing
the total fluxes $\left<dX/dt\right>$ we find that our code always
agrees with the equatorial code at the level of their claimed
fractional accuracy of $10^{-3}$.  We have tested this for the top
panel, for $a=0.5$, of Table VII of Ref.~\cite{equatorial}.  Note that
the fractional accuracy of $10^{-3}$ is the most conservative of the
accuracy claims made in Ref.~\cite{equatorial}.  Usually we find
agreement with the equatorial code well beyond this figure; for many
of the total fluxes, we agree with all six of their published digits.

While we find agreement with the equatorial code when computing the
\emph{total} fluxes $\left<dX/dt\right>$, the two codes show
significant differences when computing the horizon fluxes
$\left<dX/dt\right>^{\text{H}}$ (no such disagreement was found for
the fluxes at infinity $\left<dX/dt\right>^{\infty}$).  The
disagreement is especially strong for the energy flux at the horizon,
but it is also present to a lesser extent with the angular momentum
flux at the horizon.  For the energy fluxes at the horizon, we
typically only agree with the equatorial code to about 1-10\%, however
in one case (the orbit with $e = 0.4$ and $p = 5$ in Table VII of
Ref.~\cite{equatorial}, which has an exceptionally weak energy flux at
the horizon) our results differ fractionally by about a factor of
three, and they even have different signs.  Currently the source of
this disagreement is unknown.  We emphasize however that the total
flux of either energy of angular momentum is always dominated by the
flux at infinity, and even in the most extreme case of our
disagreements over the horizon fluxes, we still find acceptable
agreement for the total fluxes.

Hughes has shown that, in the Schwarzschild limit, under a
transformation from an equatorial to an inclined orbit, the individual
modes must transform as \cite{circular}\footnote{Here we correct two
typos in Eq.~(5.4) of Ref.~\cite{circular}.}
\begin{equation}
\frac{\left< dE/dt\right>^\star_{l(m-k)kn}(\theta_\text{inc})}{\left< dE/dt\right>^\star_{lm0n}(\theta_\text{inc}=0)}
 = \left|\mathcal{D}^l_{(m-k)m}(\theta_\text{inc}) \right|^2,
\end{equation} 
where $\mathcal{D}^l_{m'm}(\theta_\text{inc})$ is the Wigner D-function, 
as defined by Eq.~(4.256) in Ref.~\cite{Arfken}:
\begin{align}
& \mathcal{D}^l_{m'm}(\theta_\text{inc})
=  \sum_{q=\max(0,m'-m)}^{\min(l-m,l+m')} (-1)^q &
\nonumber \\
&\times \left(\cos \frac{\theta_\text{inc}}{2} \right)^{2l + m' - m - 2q} 
  \left(-\sin \frac{\theta_\text{inc}}{2} \right)^{m - m' + 2q}&
\nonumber \\
& \times \frac{ \sqrt{(l+m')! (l-m')! (l+m)! (l-m)!} }{q! (l-m-q)! (l+m'-q)! (m-m'+q)!} .&
\end{align}
For a variety of generic Schwarzschild orbits (which are both
eccentric and inclined) and for a variety of modes, we have checked
that our code satisfies this requirement to a fractional accuracy of
$10^{-5}$ or better.  We have also
confirmed that, again for a variety of generic Schwarzschild orbits,
the total fluxes of energy at both the horizon and infinity, are
independent of inclination (we have checked this up to a fractional
accuracy of $10^{-4}$). 

With the help of Norichika Sago, we have also compared the output of our
code with analytical post-Newtonian expressions (essentially
an extension of Ref.~\cite{Minoetalreview}) for the fluxes of
energy and angular momentum at infinity for orbits which are both slightly
eccentric and slightly inclined \cite{Sago in prep}.  We compared the
contributions for $l=2$, 3, and 4.  With a requested accuracy of $10^{-4}$ in our
numerical code, the two methods of calculation always agreed to $\sim 10^{-3}$ or better
(with slightly better agreement for angular momentum fluxes than for energy fluxes).
Preliminary investigations suggest that this disagreement is due to the
small-eccentricity-expansion used when deriving the post-Newtonian expressions, and is
therefore not unexpected.

\section{Results}
\label{s:Results}

In this section we discuss the results found from using the code
described in the previous section.  After describing the general
characteristics of the radiation from a typical generic orbit, we will
discuss how these characteristics vary over a catalog of orbits.

\subsection{Radial and polar voices}

\begin{figure}
\includegraphics[width = .45\textwidth]{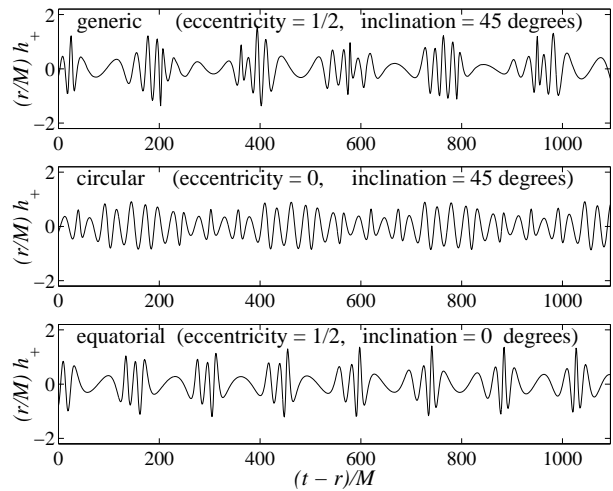}
\caption{A comparison of waveforms associated with a generic orbit and
with the related circular and equatorial orbits.  Each orbit has a
semilatus rectum of $p=4$. The magnitude of the massive black hole's
spin angular momentum is $aM=0.9M^2$. Each waveform is plotted for
$1.5 \times M/(10^6 M_\odot)$ hours as seen from a viewing angle of
$\theta = 30^\circ$ away from the symmetry axis of the massive black
hole.}
\label{f:generic}
\end{figure}

\begin{figure}
\includegraphics[width = .45\textwidth]{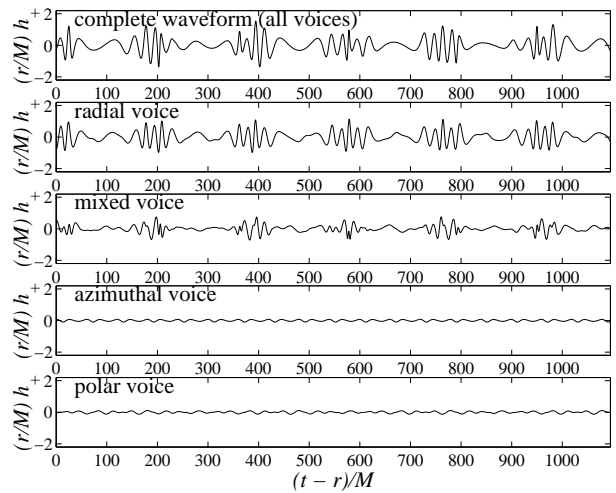}
\caption{The waveform from Fig.~\ref{f:generic} and its separated
pieces Eq.~(\ref{H pieces}).  This orbit has an eccentricity $e=0.5$,
inclination $\theta_\text{inc} = 45^\circ$, and semilatus rectum
$p=4$.  The spin of the black hole is $a=0.9M$. We plot the
waveforms as seen from a viewing angle of $\theta = 30^\circ$.  A plot
of $h_\times$ shows similar behavior.}
\label{f:H split}
\end{figure}

Figure \ref{f:generic} compares a snapshot waveform produced by a
generic orbit with the waveforms produced by the circular and
equatorial limits of that orbit.
It is clear that the generic waveform is not well approximated by
either of the limiting cases at the level of accuracy needed to
produce detection waveforms (fractional phase accuracy of $\sim
10^{-4}$).  We can however understand the generic waveform as a
composition of the effects produced by the different components of the
orbital motion: the polar motion $\theta(\lambda)$ and the radial
motion $r(\lambda)$.  To do so, we first define the general waveform
quantity
\begin{equation} \label{modes}
H = h_+ - i h_\times = \sum_{kn} H_{kn} e^{-i\omega_{kn}(t-r)},
\end{equation} 
where
\begin{align}
&H_{kn} = -\frac{2}{r}\sum_{lm} \frac{Z_{lmkn}^\text{H}}{\omega_{mkn}^{2}}
                   S_{lmkn}(\theta) e^{im[\phi-\Omega_\phi(t-r)]},& \\
&\omega_{kn} = k\Omega_\theta + n\Omega_r.&
\label{eq:Hkn_def}
\end{align}
Each term $H_{kn} \exp[-i\omega_{kn}(t-r)]$ in the sum (\ref{modes})
can be likened to a ``voice'' in the GW produced by the orbit.  By
using the general expression for the $Z^\star_{lmkn}$ coefficients
(\ref{final Z}) and the symmetry rules (\ref{geodesic sym}) for the
geodesic expansions, we see that in the limiting circular and
equatorial cases, only some voices contribute
\begin{subequations}\label{waveform sym}
\begin{align}
H_{kn} &\propto \delta_{n0},& &\text{(circular orbits)}&
\\
H_{kn} &\propto \delta_{k0},& &\text{(equatorial orbits)}&
\\
H_{kn} &\propto \delta_{n0}\delta_{k0}.& &\text{(circular-equatorial orbits)}&
\end{align}
\end{subequations}
The term $H_{00}$, the sum of all the terms which oscillate at pure
multiples of the azimuthal frequency, is the only nonvanishing term
for orbits which are both circular and equatorial.  The polar and
radial voices are made of terms which oscillate at integer linear
combinations of the azimuthal frequency, and either the radial or
polar frequency.  The polar voices, with $n=0$ and $k \ne 0$, can be
thought of as being produced by the orbital inclination.  Similarly,
the radial voices ($k=0$ and $n \ne 0$) are produced by the orbit's
eccentricity.  This suggests the following separation of the voices:
\begin{equation} \label{H split}
H = H_\text{azimuthal} + H_\text{polar} + H_\text{radial} + H_\text{mixed},
\end{equation}
where
\begin{subequations}
\begin{align} \label{H pieces}
H_\text{azimuthal}  &= H_{00},&
\\
H_\text{polar}  &= \sum_{k\ne0} H_{k0} e^{-i\omega_{k0}t},&
\\
H_\text{radial} &= \sum_{n\ne0} H_{0n} e^{-i\omega_{0n}t},&
\\
H_\text{mixed}  &= \sum_{k\ne0} \sum_{n\ne0} H_{kn} e^{-i\omega_{kn}t}.&
\end{align}
\end{subequations}
Figure \ref{f:H split} displays the waveforms for the different voices
of the generic waveform from Fig.~\ref{f:generic}.
Since that orbit has eccentricity $e=0.5$ and inclination
$\theta_\text{inc} = 45^\circ$, one might expect that the radial and
polar voices would be comparably loud. Instead (as one might guess by
looking at the waveform from the equatorial limit of this orbit; cf.\
Fig.~\ref{f:generic}), the radial voice alone $H_\text{radial}$ has a
phase history which is similar to that of the complete waveform.  The
polar voice $H_\text{polar}$ is comparatively unimportant.

The fluxes in energy and angular momentum can be similarly separated
into radial, polar, azimuthal, and mixed pieces.  Again for this case,
the radial voice dominates.  Table \ref{t:split fluxes} shows how the
fluxes are distributed in the case of the orbit used to produce
Figs.~\ref{f:generic} and \ref{f:H split}.
\begin{table}
\begin{ruledtabular}
\begin{tabular}{c|cccc}
                       & radial & polar  & azimuthal & mixed \\ \hline
$\left<dE/dt\right>$   &  62\%  & 1.2\%  & 0.52\%    & 37\%  \\ \hline
$\left<dL_z/dt\right>$ &  73\%  & 0.77\% & 1.0\%     & 25\% 
\end{tabular}
\end{ruledtabular}
\caption{\label{t:split fluxes} The results of applying the splitting procedure
described by Eq.~(\ref{H split}) to the fluxes of energy $E$ and angular momentum $L_z$ produced by 
the generic orbit from Fig.~\ref{f:generic}. Each number is the ratio of the contribution from a specific voice,
to the total quantity.}
\end{table}
The radial voice carries away more than half of both the energy and
angular momentum fluxes, while the polar voice carries only about
1\%. This suggests that GWs from this orbit might, for some
applications, be well approximated by its radial voice alone: $H
\approx H_\text{radial}$.  Such an approximation would by highly
desirable since it would considerably reduce the computational cost of
generating this waveform.  Unfortunately $H_\text{radial}$ is still
quite sensitive to the orbital inclination, as can be seen by
comparing to the waveforms for the circular and equatorial orbits in
Fig.~\ref{f:generic}.  This is because the inclination influences the
values of all three orbital frequencies. So even if this approximation
were valid for some range of orbits, it would not likely reduce the
number of detection waveforms needed when searching for events
produced by those orbits, though it may reduce the computational cost
to produce each waveform.

\subsection{Catalog of orbits}

We now discuss the waveforms and fluxes for a catalog of generic orbits. 
The parameters of these orbits are shown in Table \ref{t:orbits}.
\begin{table*}
\begin{ruledtabular}
\begin{tabular}{ccc||cccc}
$e$ &  
$p$ &
$\theta_\text{inc}$ &
$E/\mu$ &
$L_z/(\mu M)$ &
$Q / (\mu^2 M^2)$ \\
\hline
0.1 &  6 &  20$^\circ$ & 0.923976142907 &  2.65115182126    &  0.944969071904\\
0.1 &  6 &  40$^\circ$ & 0.926051429284 &  2.22165013064    &  3.52285570065 \\
0.1 &  6 &  60$^\circ$ & 0.929863620985 &  1.52003893228    &  7.01378240707 \\
0.1 &  6 &  80$^\circ$ & 0.935903446174 &  0.564178168635   & 10.3350037833  \\
0.1 & 12 & 100$^\circ$ & 0.962956523141 & $-$0.701069332316 & 15.8653948211  \\
0.1 & 12 & 120$^\circ$ & 0.963964137935 & $-$2.07109099890  & 12.9112484603  \\
0.1 & 12 & 140$^\circ$ & 0.964900141027 & $-$3.24526311278  &  7.43835020273 \\
0.1 & 12 & 160$^\circ$ & 0.965574148307 & $-$4.04295255570  &  2.17176613884 \\
\hline
0.3 &  6 &  20$^\circ$ & 0.929683266598 &  2.66499623815    &  0.953716658814\\
0.3 &  6 &  40$^\circ$ & 0.931463021604 &  2.23504604006    &  3.56152665553 \\
0.3 &  6 &  60$^\circ$ & 0.934742155317 &  1.53133514084    &  7.11166313087 \\
0.3 &  6 &  80$^\circ$ & 0.939961770755 &  0.569540184106   & 10.5245453291  \\
0.3 & 12 & 100$^\circ$ & 0.965665301518 & $-$0.704444574360 & 16.0138725377  \\
0.3 & 12 & 120$^\circ$ & 0.966529280028 & $-$2.08286664248  & 13.0549867002  \\
0.3 & 12 & 140$^\circ$ & 0.967333822046 & $-$3.26633491912  &  7.53338516325 \\
0.3 & 12 & 160$^\circ$ & 0.967914357052 & $-$4.07155287011  &  2.20208170741 \\
\hline
0.5 &  6 &  20$^\circ$ & 0.941309757948 &  2.69327096428    &  0.971725840431\\
0.5 &  6 &  40$^\circ$ & 0.942555891555 &  2.26253463210    &  3.64161734490 \\
0.5 &  6 &  60$^\circ$ & 0.944866295222 &  1.55467904187    &  7.31622158779 \\
0.5 &  6 &  80$^\circ$ & 0.948580242140 &  0.580718189662   & 10.9253034674  \\
0.5 & 12 & 100$^\circ$ & 0.971221837220 & $-$0.711332909044 & 16.3190869455  \\
0.5 & 12 & 120$^\circ$ & 0.971826517606 & $-$2.10697070044  & 13.3517251785  \\
0.5 & 12 & 140$^\circ$ & 0.972392433557 & $-$3.30957999464  &  7.73032698828 \\
0.5 & 12 & 160$^\circ$ & 0.972802509120 & $-$4.13035470435  &  2.26507351639 \\
\hline
0.7 &  6 &  20$^\circ$ & 0.959306993445 &  2.73722358422    &  1.00010433263 \\
0.7 &  6 &  40$^\circ$ & 0.959910828459 &  2.30561707530    &  3.76913698051 \\
0.7 &  6 &  60$^\circ$ & 0.961041797423 &  1.59171983094    &  7.64712824881 \\
0.7 &  6 &  80$^\circ$ & 0.962889252972 &  0.598733562204   & 11.5872341739  \\
0.7 & 12 & 100$^\circ$ & 0.979920978103 & $-$0.722026359625 & 16.7987400685  \\
0.7 & 12 & 120$^\circ$ & 0.980213741124 & $-$2.14458473041  & 13.8215334684  \\
0.7 & 12 & 140$^\circ$ & 0.980489898895 & $-$3.37737131644  &  8.04421001469 \\
0.7 & 12 & 160$^\circ$ & 0.980691338739 & $-$4.22282649751  &  2.36594094475 
\end{tabular}         
\caption{\label{t:orbits}
Each row above describes an orbit in our catalog.  The numbers to the right hand
side of the vertical bar were computed (to an fractional accuracy of $10^{-12}$)
from the numbers on the numbers on the left hand of the bar (using the 
method described in Appendix \ref{s:Geodesic details}).
Each orbit has $a = 0.9M$.
}
\end{ruledtabular}
\end{table*}
Tables \ref{t:voices_P} and \ref{t:voices_R} show the relative
contributions of the different voices for these waveforms.
\begin{table*}
\begin{ruledtabular}
\begin{tabular}{cc||cccc|cccc} 
$e$ &
$\theta_\text{inc}$ &
$\mathcal{P}_\text{radial}$     &
$\mathcal{P}_\text{polar}$     &
$\mathcal{P}_\text{azimuthal}$  & 
$\mathcal{P}_\text{mixed}$  & 
$\mathcal{T}_{z\text{,radial}}$     &
$\mathcal{T}_{z\text{,polar}}$      &
$\mathcal{T}_{z\text{,azimuthal}}$ &
$\mathcal{T}_{z\text{,mixed}}$ \\
\hline 
0.1 & $20^\circ$ & $16$\% & $8.3$\% & $74$\% & $1.8$\% & $14$\% & $ 5$\% & $80$\% & $ 1$\% \\  
0.1 & $40^\circ$ & $13$\% & $28$\% & $52$\% & $6.6$\% & $14$\% & $17$\% & $65$\% & $ 4$\% \\  
0.1 & $60^\circ$ & $9.4$\% & $48$\% & $29$\% & $14$\% & $14$\% & $29$\% & $48$\% & $9.5$\% \\  
0.1 & $80^\circ$ & $6.7$\% & $53$\% & $10$\% & $30$\% & $20$\% & $24$\% & $33$\% & $23$\% \\  
\hline 
0.3 & $20^\circ$ & $76$\% & $1.9$\% & $14$\% & $7.9$\% & $75$\% & $1.3$\% & $19$\% & $4.7$\% \\  
0.3 & $40^\circ$ & $58$\% & $5.4$\% & $7.8$\% & $29$\% & $67$\% & $3.6$\% & $12$\% & $17$\% \\  
0.3 & $60^\circ$ & $37$\% & $5.9$\% & $1.8$\% & $55$\% & $58$\% & $2.7$\% & $3.8$\% & $36$\% \\  
0.3 & $80^\circ$ & $17$\% & $1.9$\% & $0.48$\% & $80$\% & $51$\% & $-2.4$\% & $1.9$\% & $49$\% \\  
\hline 
0.5 & $20^\circ$ & $90$\% & $0.077$\% & $0.46$\% & $9.5$\% & $93$\% & $0.083$\% & $0.92$\% & $ 6$\% \\  
0.5 & $40^\circ$ & $66$\% & $0.24$\% & $0.64$\% & $33$\% & $77$\% & $0.18$\% & $1.5$\% & $21$\% \\  
0.5 & $60^\circ$ & $39$\% & $0.93$\% & $1.1$\% & $59$\% & $58$\% & $0.68$\% & $3.3$\% & $38$\% \\  
0.5 & $80^\circ$ & $19$\% & $1.6$\% & $0.0055$\% & $80$\% & $53$\% & $-2$\% & $0.029$\% & $49$\% \\  
\hline 
0.7 & $20^\circ$ & $90$\% & $0.057$\% & $0.62$\% & $9.3$\% & $92$\% & $0.1$\% & $2.3$\% & $ 6$\% \\  
0.7 & $40^\circ$ & $67$\% & $0.22$\% & $0.36$\% & $33$\% & $77$\% & $0.35$\% & $1.5$\% & $21$\% \\  
0.7 & $60^\circ$ & $40$\% & $0.19$\% & $0.0018$\% & $59$\% & $61$\% & $-0.17$\% & $0.0094$\% & $39$\% \\  
0.7 & $80^\circ$ & $19$\% & $0.16$\% & $0.043$\% & $81$\% & $53$\% & $-0.47$\% & $0.33$\% & $47$\%   
\end{tabular}
\caption{\label{t:voices_P} Here we show the results of applying the splitting procedure
described by Eq.~(\ref{H split}) to the fluxes of energy $E$ and angular momentum $L_z$ produced by our 
catalog of generic orbits (Table \ref{t:orbits}).  The symbol $\mathcal{P}$ represents 
a ratio of the average power relative to the total average power $\left< dE/dt\right>$, and the symbol 
$\mathcal{T}$ represents a ratio of the average torque to the total average torque $\left< dL_z/dt\right>$.
Each of these orbits has $a = 0.9M$ and $p = 6$.}
\end{ruledtabular}
\end{table*}
\begin{table*}
\begin{ruledtabular}
\begin{tabular}{cc||cccc|cccc} 
$e$ &
$\theta_\text{inc}$ &
$\mathcal{P}_\text{radial}$     &
$\mathcal{P}_\text{polar}$     &
$\mathcal{P}_\text{azimuthal}$  & 
$\mathcal{P}_\text{mixed}$  & 
$\mathcal{T}_{z\text{,radial}}$     &
$\mathcal{T}_{z\text{,polar}}$      &
$\mathcal{T}_{z\text{,azimuthal}}$ &
$\mathcal{T}_{z\text{,mixed}}$ \\
\hline
0.1 & $100^\circ$ & $1.1$\% & $76$\% & $7.7$\% & $15$\% & $7.8$\% & $23$\% & $67$\% & $1.3$\% \\
0.1 & $120^\circ$ & $ 4$\% & $60$\% & $22$\% & $14$\% & $7.4$\% & $35$\% & $51$\% & $6.5$\% \\
0.1 & $140^\circ$ & $10$\% & $35$\% & $45$\% & $9.6$\% & $12$\% & $20$\% & $63$\% & $4.8$\% \\
0.1 & $160^\circ$ & $18$\% & $11$\% & $68$\% & $3.3$\% & $17$\% & $5.7$\% & $76$\% & $1.6$\% \\
\hline
0.3 & $100^\circ$ & $6.1$\% & $17$\% & $2.1$\% & $75$\% & $52$\% & $13$\% & $26$\% & $9.4$\% \\
0.3 & $120^\circ$ & $20$\% & $10$\% & $4.8$\% & $65$\% & $43$\% & $8.4$\% & $14$\% & $34$\% \\
0.3 & $140^\circ$ & $47$\% & $4.3$\% & $7.1$\% & $42$\% & $62$\% & $3.2$\% & $13$\% & $22$\% \\
0.3 & $160^\circ$ & $78$\% & $0.94$\% & $7.7$\% & $13$\% & $81$\% & $0.65$\% & $11$\% & $6.9$\% \\
\hline
0.5 & $100^\circ$ & $7.5$\% & $0.26$\% & $0.042$\% & $92$\% & $80$\% & $1.2$\% & $0.88$\% & $18$\% \\
0.5 & $120^\circ$ & $23$\% & $0.37$\% & $0.038$\% & $76$\% & $56$\% & $0.5$\% & $0.18$\% & $43$\% \\
0.5 & $140^\circ$ & $52$\% & $0.51$\% & $0.34$\% & $47$\% & $72$\% & $0.59$\% & $0.97$\% & $26$\% \\
0.5 & $160^\circ$ & $84$\% & $0.25$\% & $1.2$\% & $15$\% & $89$\% & $0.27$\% & $2.6$\% & $7.8$\% \\
\hline
0.7 & $100^\circ$ & $6.8$\% & $0.36$\% & $0.038$\% & $93$\% & $84$\% & $0.66$\% & $1.8$\% & $13$\% \\
0.7 & $120^\circ$ & $22$\% & $0.32$\% & $0.15$\% & $78$\% & $53$\% & $ 1$\% & $1.5$\% & $44$\% \\
0.7 & $140^\circ$ & $50$\% & $0.16$\% & $0.28$\% & $49$\% & $70$\% & $0.4$\% & $1.5$\% & $28$\% \\
0.7 & $160^\circ$ & $84$\% & $0.019$\% & $0.22$\% & $16$\% & $91$\% & $0.043$\% & $0.92$\% & $8.5$\%
\end{tabular}
\caption{\label{t:voices_R} Identical 
to Table \ref{t:voices_P}, except that these orbits 
are retrograde and have semilatus rectum $p = 12$.}
\end{ruledtabular}
\end{table*}

It is clear that as inclination is increased, radiation is channeled
into the polar voice, and that likewise the radial voice is amplified
by an increase in eccentricity.  Although both voices can be amplified
by such adjustments, the radial voice seems particularly booming while
the polar voice is more subtle.  While the polar voice can be made
dominant, it requires especially low eccentricity, below about 0.3.
Also, for orbits with eccentricity as low as 0.1 and inclinations 
$\theta_\text{inc} \lesssim 45^\circ$ or $\theta_\text{inc} \gtrsim 135^\circ$,
one should expect to capture at least half of the radiative power and torque in the azimuthal voice alone.  This is especially significant since the azimuthal voice is \emph{exceedingly} inexpensive to compute in comparison to the others. 
It is composed only of oscillations at pure multiples of the azimuthal frequency with $k = n = 0$, so that computing it alone would mean replacing
all of the  four dimensional sums in Sec.~\ref{ss:Waveforms and fluxes} with only two dimensional sums.

The asymptotic energy and angular momentum fluxes associated with the
catalog (Table \ref{t:orbits}) are displayed in Tables
\ref{t:fluxes_P} and \ref{t:fluxes_R}.  The energy and angular
momentum fluxes were requested to have fractional accuracies of
$\varepsilon_\text{flux} = 10^{-4}$.  The estimates of the actual
fractional accuracies, shown in square brackets in Tables
\ref{t:fluxes_P} and \ref{t:fluxes_R}, were determined as follows:
First we compute each flux with a requested fractional accuracy of
$\varepsilon_\text{flux} = 10^{-4}$.  The fractional accuracy claimed
in Tables \ref{t:fluxes_P} and \ref{t:fluxes_R} is then either
$10^{-4}$, or if larger, the fractional residual that was found when comparing with
the same flux computed with a requested fractional accuracy of
$\varepsilon_\text{flux} = 10^{-3}$.  The summation buffer $B$ for the
polar harmonic index $k$ is 3, while $B=5$ for the radial harmonic
index $n$.  The largest $l$ value needed for the $e=0.1,~0.3$ orbits
was $l=10$, while for the $e=0.5,~0.7$ orbits the largest $l$ value was
$l=11$. The most complex waveform,
$(e,\theta_\text{inc})=(0.7,80^\circ)$, is made up of about $160,000$
modes (about 80,000 mode calculations due to symmetry). The simplest
waveform, $(e,\theta_\text{inc})=(0.1,20^\circ)$, is made from about
17,000 modes (about 8,500 mode calculations).  
\begin{table*}
\begin{ruledtabular}
\begin{tabular}{cc||ccccc}
$e$ &  
$\theta_\text{inc}$ &  
$\left< dE/dt \right>^\text{H} (M/\mu)^2$ &
$\left< dE/dt \right>^\infty (M/\mu)^2$ &
$\left< dL_z/dt \right>^\text{H} M/ \mu^2$ &
$\left< dL_z/dt \right>^\infty M/ \mu^2$ &
$\left< dQ/dt \right> / (\mu^2 M)$ \\
\hline
0.1 & $20^\circ$ & $-4.25756\times 10^{-6} [10^{-4}]$ & $5.87399\times 10^{-4} [10^{-4}]$ & $-6.72410\times 10^{-5} [10^{-4}]$ & $8.53652\times 10^{-3} [10^{-4}]$ & $6.03753\times 10^{-3}$ \\
0.1 & $40^\circ$ & $-3.96692\times 10^{-6} [10^{-4}]$ & $6.18500\times 10^{-4} [10^{-4}]$ & $-7.76672\times 10^{-5} [10^{-4}]$ & $7.62823\times 10^{-3} [10^{-4}]$ & $2.39458\times 10^{-2}$ \\
0.1 & $60^\circ$ & $-3.36171\times 10^{-6} [10^{-4}]$ & $6.83855\times 10^{-4} [10^{-4}]$ & $-1.12143\times 10^{-4} [10^{-4}]$ & $6.07264\times 10^{-3} [10^{-4}]$ & $5.50060\times 10^{-2}$ \\
0.1 & $80^\circ$ & $-9.78653\times 10^{-7} [10^{-4}]$ & $8.07007\times 10^{-4} [10^{-4}]$ & $-1.90843\times 10^{-4} [10^{-4}]$ & $3.61820\times 10^{-3} [10^{-4}]$ & $1.25570\times 10^{-1}$ \\
\hline
0.3 & $20^\circ$ & $-5.88185\times 10^{-6} [10^{-4}]$ & $6.80432\times 10^{-4} [10^{-4}]$ & $-7.78419\times 10^{-5} [10^{-4}]$ & $8.62437\times 10^{-3} [10^{-4}]$ & $6.11706\times 10^{-3}$ \\
0.3 & $40^\circ$ & $-5.88820\times 10^{-6} [10^{-4}]$ & $7.26781\times 10^{-4} [10^{-4}]$ & $-1.00628\times 10^{-4} [10^{-4}]$ & $7.83499\times 10^{-3} [10^{-4}]$ & $2.46493\times 10^{-2}$ \\
0.3 & $60^\circ$ & $-5.28978\times 10^{-6} [10^{-4}]$ & $8.31504\times 10^{-4} [10^{-4}]$ & $-1.66905\times 10^{-4} [10^{-4}]$ & $6.48662\times 10^{-3} [10^{-4}]$ & $5.86987\times 10^{-2}$ \\
0.3 & $80^\circ$ & $-1.52779\times 10^{-7} [10^{-4}]$ & $1.08629\times 10^{-3} [10^{-4}]$ & $-3.46171\times 10^{-4} [10^{-4}]$ & $4.36910\times 10^{-3} [10^{-4}]$ & $1.48680\times 10^{-1}$ \\
\hline
0.5 & $20^\circ$ & $-8.37384\times 10^{-6} [10^{-4}]$ & $7.98857\times 10^{-4} [10^{-4}]$ & $-9.16902\times 10^{-5} [10^{-4}]$ & $8.34425\times 10^{-3} [10^{-4}]$ & $5.95501\times 10^{-3}$ \\
0.5 & $40^\circ$ & $-9.06408\times 10^{-6} [10^{-4}]$ & $8.74449\times 10^{-4} [10^{-3}]$ & $-1.38234\times 10^{-4} [10^{-4}]$ & $7.80844\times 10^{-3} [10^{-3}]$ & $2.46909\times 10^{-2}$ \\
0.5 & $60^\circ$ & $-8.31005\times 10^{-6} [10^{-4}]$ & $1.05986\times 10^{-3} [10^{-3}]$ & $-2.70993\times 10^{-4} [10^{-4}]$ & $6.92901\times 10^{-3} [10^{-3}]$ & $6.26645\times 10^{-2}$ \\
0.5 & $80^\circ$ & $ 5.67035\times 10^{-6} [10^{-4}]$ & $1.67918\times 10^{-3} [10^{-3}]$ & $-7.40721\times 10^{-4} [10^{-4}]$ & $5.87398\times 10^{-3} [10^{-3}]$ & $1.93149\times 10^{-1}$ \\
\hline
0.7 & $20^\circ$ & $-9.34196\times 10^{-6} [10^{-4}]$ & $7.69363\times 10^{-4} [10^{-2}]$ & $-8.95146\times 10^{-5} [10^{-4}]$ & $6.66378\times 10^{-3} [10^{-2}]$ & $4.80410\times 10^{-3}$ \\
0.7 & $40^\circ$ & $-1.07089\times 10^{-5} [10^{-3}]$ & $8.70367\times 10^{-4} [10^{-2}]$ & $-1.57713\times 10^{-4} [10^{-4}]$ & $6.48763\times 10^{-3} [10^{-2}]$ & $2.06958\times 10^{-2}$ \\
0.7 & $60^\circ$ & $-9.23969\times 10^{-6} [10^{-3}]$ & $1.13494\times 10^{-3} [10^{-2}]$ & $-3.62611\times 10^{-4} [10^{-4}]$ & $6.29440\times 10^{-3} [10^{-2}]$ & $5.69964\times 10^{-2}$ \\
0.7 & $80^\circ$ & $ 2.85367\times 10^{-5} [10^{-1}]$ & $2.69226\times 10^{-3} [10^{-1}]$ & $-1.58165\times 10^{-3} [10^{-2}]$ & $8.25056\times 10^{-3} [10^{-1}]$ & $2.58125\times 10^{-1}$ 
\end{tabular}                                                                                                                           
\caption{\label{t:fluxes_P} The fluxes of energy $E$ and axial angular momentum $L_z$ at infinity                                
(superscript of $\infty$) and through the black hole's event horizon (superscript of H) for some generic orbits.
Each of these orbits has $a = 0.9M$ and $p = 6$.  Each number
in square brackets is an order of magnitude estimate for the fractional accuracy of the preceding number. 
The rates of change for the Carter constant $\left< dQ/dt \right>$ were computed from the formula (\ref{Qdot}) which follows
from the assumption that radiation does not change the orbital inclination.
Since this is an uncontrolled approximation, the accuracy of the $\left< dQ/dt \right>$
figures is unknown; we include these data for comparison purposes.
}                                                                                           
\end{ruledtabular}
\end{table*}
\begin{table*}
\begin{ruledtabular}
\begin{tabular}{cc||ccccc}
$e$ &  
$\theta_\text{inc}$ &  
$\left< dE/dt \right>^\text{H} (M/\mu)^2$ &
$\left< dE/dt \right>^\infty (M/\mu)^2$ &
$\left< dL_z/dt \right>^\text{H} M/ \mu^2$ &
$\left< dL_z/dt \right>^\infty M/ \mu^2$ &
$\left< dQ/dt \right> / (\mu^2 M)$ \\
\hline
0.1 & $100^\circ$ & $1.10409\times 10^{-8} [10^{-4}]$ & $2.50845\times 10^{-5} [10^{-3}]$ & $-1.85760\times 10^{-6} [10^{-4}]$ & $-1.22779\times 10^{-4} [10^{-2}]$ & $5.64112\times 10^{-3}$ \\
0.1 & $120^\circ$ & $3.80216\times 10^{-8} [10^{-4}]$ & $2.67639\times 10^{-5} [10^{-4}]$ & $-2.47774\times 10^{-6} [10^{-4}]$ & $-4.91749\times 10^{-4} [10^{-4}]$ & $6.16205\times 10^{-3}$ \\
0.1 & $140^\circ$ & $7.36704\times 10^{-8} [10^{-4}]$ & $2.83939\times 10^{-5} [10^{-4}]$ & $-3.36034\times 10^{-6} [10^{-4}]$ & $-8.40626\times 10^{-4} [10^{-4}]$ & $3.86894\times 10^{-3}$ \\
0.1 & $160^\circ$ & $1.06500\times 10^{-7} [10^{-4}]$ & $2.96240\times 10^{-5} [10^{-4}]$ & $-4.23915\times 10^{-6} [10^{-4}]$ & $-1.09777\times 10^{-3} [10^{-4}]$ & $1.18394\times 10^{-3}$ \\
\hline
0.3 & $100^\circ$ & $2.44837\times 10^{-8} [10^{-4}]$ & $2.92408\times 10^{-5} [10^{-4}]$ & $-2.74782\times 10^{-6} [10^{-4}]$ & $-1.14269\times 10^{-4} [10^{-4}]$ & $5.32021\times 10^{-3}$ \\
0.3 & $120^\circ$ & $7.99022\times 10^{-8} [10^{-4}]$ & $3.18207\times 10^{-5} [10^{-4}]$ & $-3.86655\times 10^{-6} [10^{-4}]$ & $-4.92316\times 10^{-4} [10^{-4}]$ & $6.21994\times 10^{-3}$ \\
0.3 & $140^\circ$ & $1.57246\times 10^{-7} [10^{-4}]$ & $3.45503\times 10^{-5} [10^{-4}]$ & $-5.43045\times 10^{-6} [10^{-4}]$ & $-8.65358\times 10^{-4} [10^{-4}]$ & $4.01673\times 10^{-3}$ \\
0.3 & $160^\circ$ & $2.32108\times 10^{-7} [10^{-4}]$ & $3.67886\times 10^{-5} [10^{-4}]$ & $-6.98894\times 10^{-6} [10^{-4}]$ & $-1.15339\times 10^{-3} [10^{-4}]$ & $1.25517\times 10^{-3}$ \\
\hline
0.5 & $100^\circ$ & $6.01938\times 10^{-8} [10^{-4}]$ & $3.44049\times 10^{-5} [10^{-4}]$ & $-4.38711\times 10^{-6} [10^{-4}]$ & $-9.51540\times 10^{-5} [10^{-4}]$ & $4.56726\times 10^{-3}$ \\
0.5 & $120^\circ$ & $1.87174\times 10^{-7} [10^{-4}]$ & $3.87672\times 10^{-5} [10^{-4}]$ & $-6.64495\times 10^{-6} [10^{-4}]$ & $-4.68884\times 10^{-4} [10^{-4}]$ & $6.02679\times 10^{-3}$ \\
0.5 & $140^\circ$ & $3.77537\times 10^{-7} [10^{-4}]$ & $4.37778\times 10^{-5} [10^{-4}]$ & $-9.86398\times 10^{-6} [10^{-4}]$ & $-8.67384\times 10^{-4} [10^{-4}]$ & $4.09805\times 10^{-3}$ \\
0.5 & $160^\circ$ & $5.75306\times 10^{-7} [10^{-4}]$ & $4.82937\times 10^{-5} [10^{-4}]$ & $-1.31850\times 10^{-5} [10^{-4}]$ & $-1.20168\times 10^{-3} [10^{-4}]$ & $1.33246\times 10^{-3}$ \\
\hline
0.7 & $100^\circ$ & $1.16031\times 10^{-7} [10^{-3}]$ & $3.29140\times 10^{-5} [10^{-4}]$ & $-5.69490\times 10^{-6} [10^{-4}]$ & $-6.19594\times 10^{-5} [10^{-3}]$ & $3.14810\times 10^{-3}$ \\
0.7 & $120^\circ$ & $3.48444\times 10^{-7} [10^{-3}]$ & $3.87999\times 10^{-5} [10^{-2}]$ & $-9.38008\times 10^{-6} [10^{-3}]$ & $-3.67910\times 10^{-4} [10^{-2}]$ & $4.86316\times 10^{-3}$ \\
0.7 & $140^\circ$ & $7.27850\times 10^{-7} [10^{-4}]$ & $4.62297\times 10^{-5} [10^{-2}]$ & $-1.49671\times 10^{-5} [10^{-4}]$ & $-7.28739\times 10^{-4} [10^{-3}]$ & $3.54271\times 10^{-3}$ \\
0.7 & $160^\circ$ & $1.16315\times 10^{-6} [10^{-3}]$ & $5.37734\times 10^{-5} [10^{-2}]$ & $-2.12322\times 10^{-5} [10^{-3}]$ & $-1.06961\times 10^{-3} [10^{-2}]$ & $1.22235\times 10^{-3}$ \\
\end{tabular}                                                                                                                           
\caption{\label{t:fluxes_R} Identical 
to Table \ref{t:fluxes_P}, except that these orbits 
are retrograde and have semilatus rectum $p = 12$.}                                                                                           
\end{ruledtabular}
\end{table*}

\begin{figure*}
\includegraphics[width = .45\textwidth]{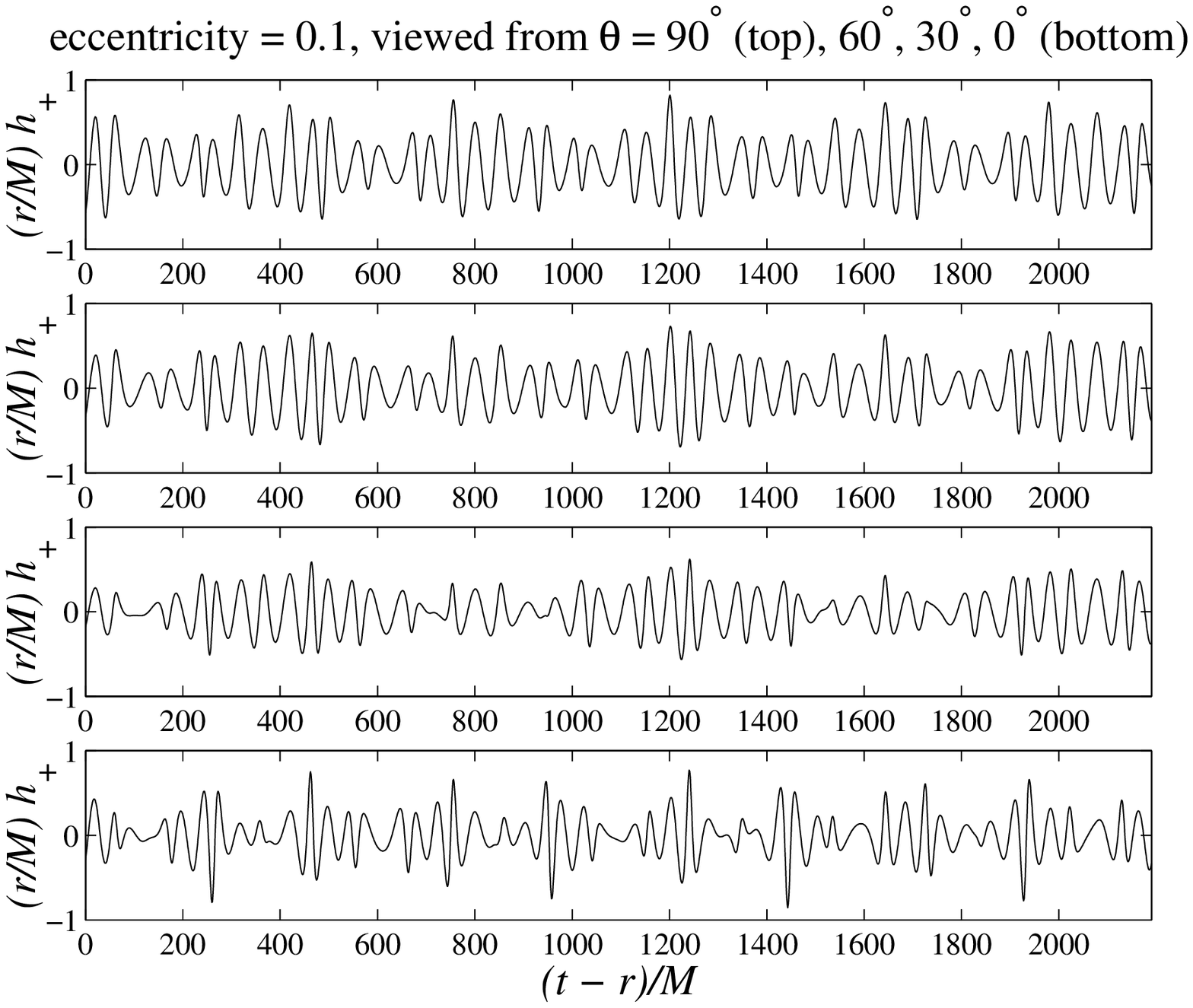}
\includegraphics[width = .45\textwidth]{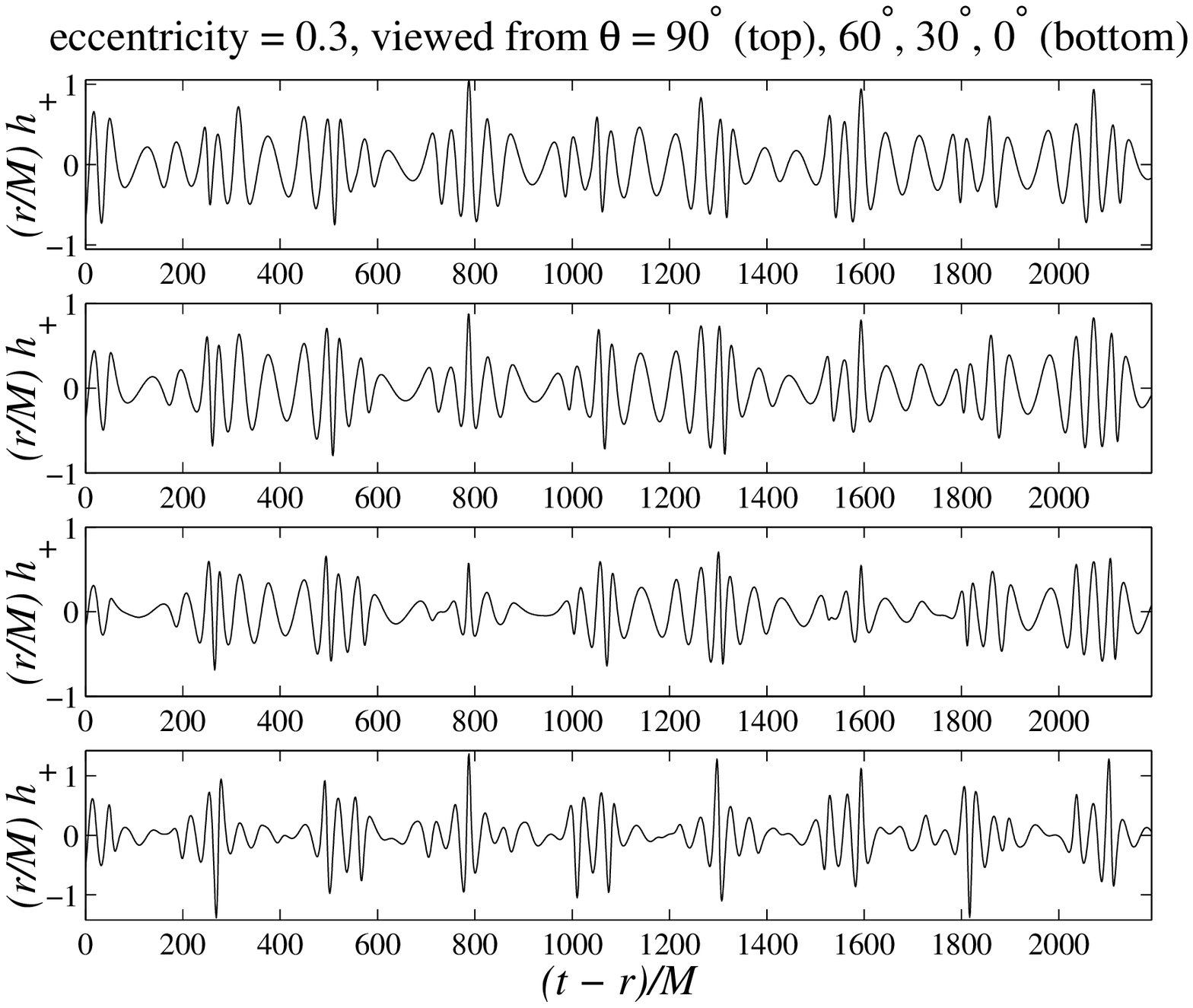}
\includegraphics[width = .45\textwidth]{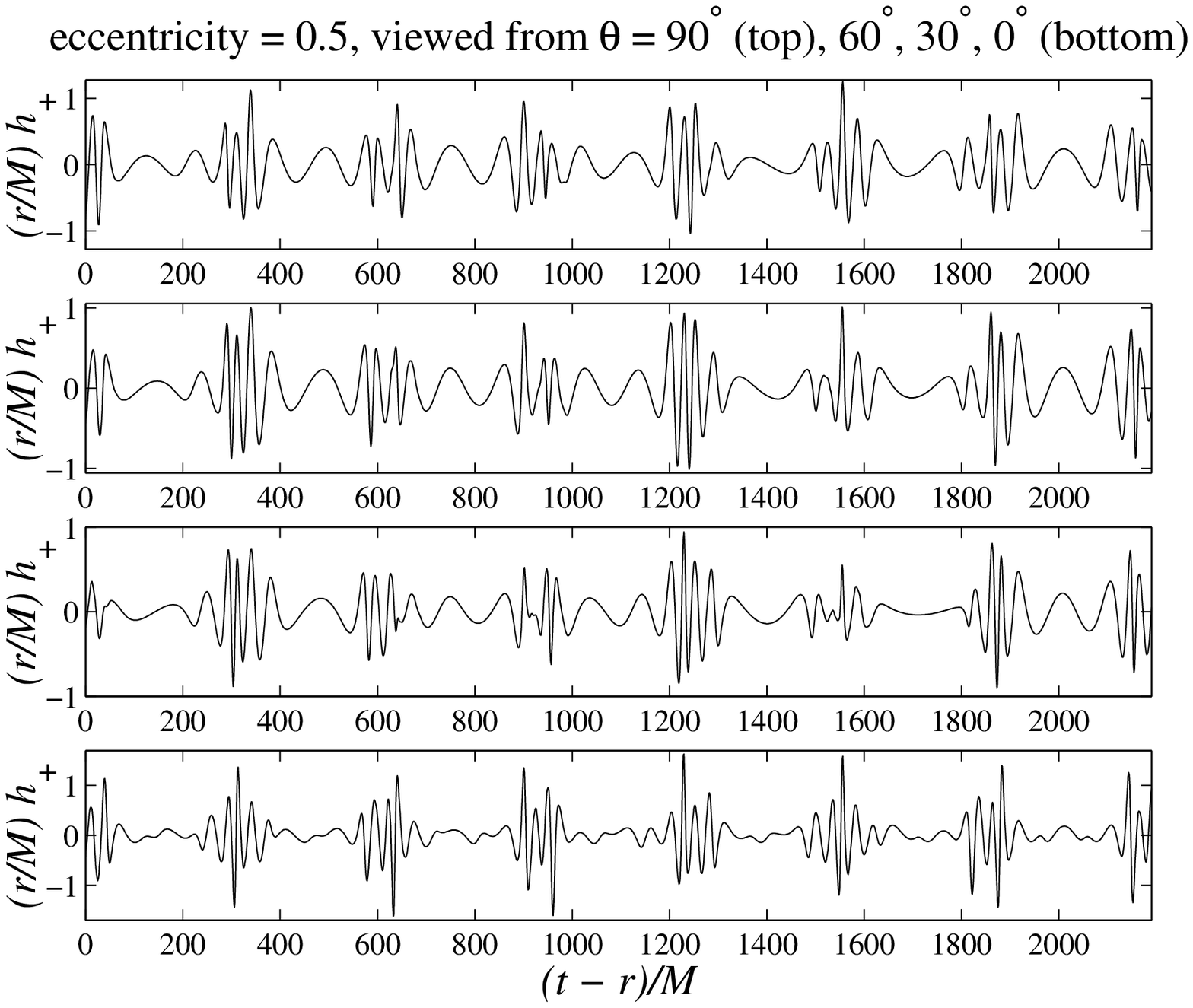}
\includegraphics[width = .45\textwidth]{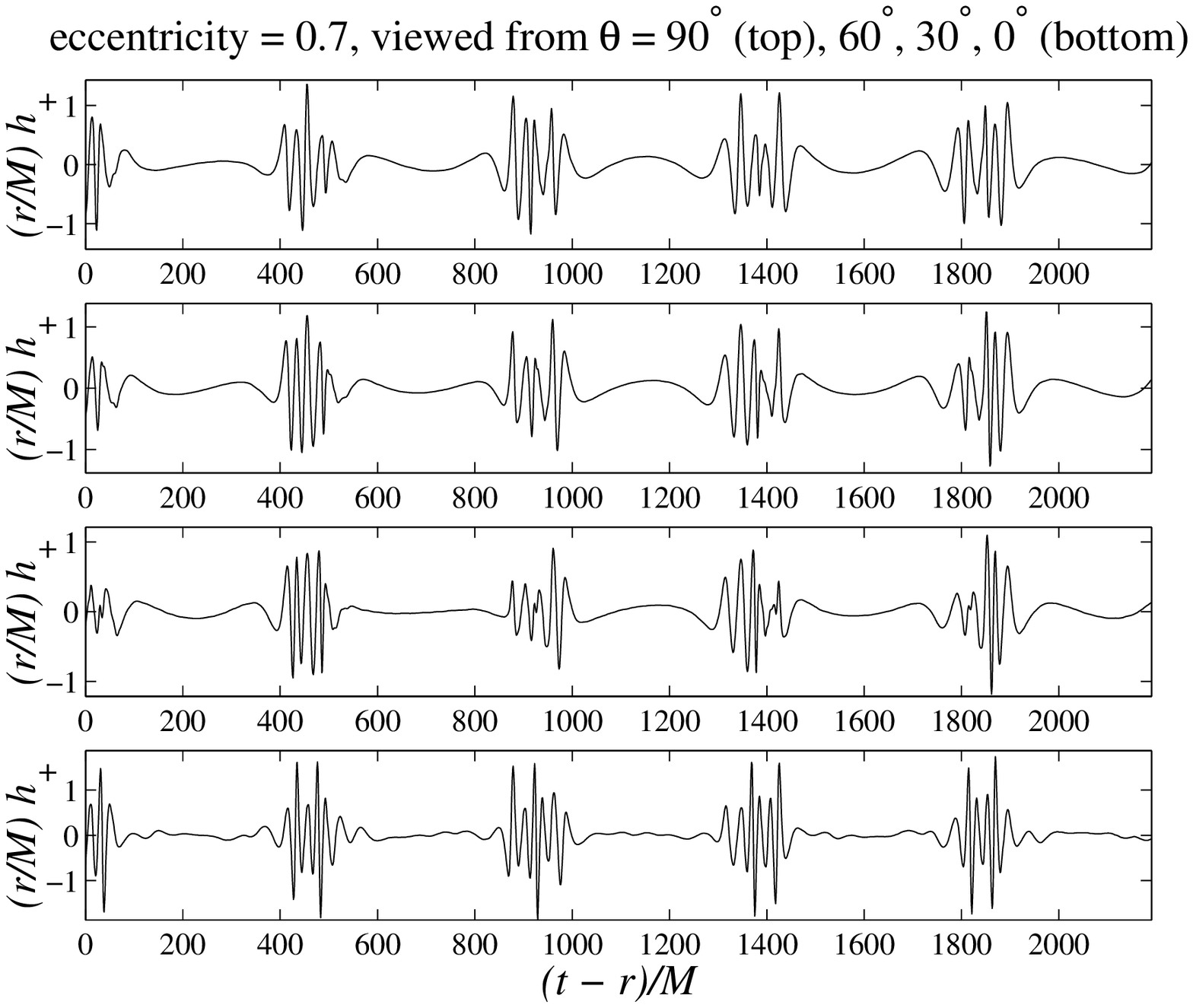}
\caption{Snapshot waveforms for orbits with inclination
$\theta_\text{inc} = 80^\circ$, semilatus rectum $p=6$, and
eccentricities $e=$ 0.1, 0.3, 0.5, 0.7. The magnitude of the black
hole's spin angular momentum is $aM=0.9M^2$.}
\label{f:80deg}
\end{figure*}

Some examples of waveforms are finally shown in the time domain in
Figs.~\ref{f:80deg}.  There we plot $h_+$ waveforms for $3 \times
M/(10^6 M_\odot)$ hours as seen from four different viewing angles
$\theta = 0^\circ,~30^\circ,~60^\circ,~90^\circ$.  These plots show
that the radial voice has a burst-like character, while the polar
voice is more of a modulated hum.  These features are consistent with
the characteristics found in earlier work which focused on circular
orbits \cite{circular} and on equatorial orbits \cite{equatorial}.

\section{Summary and future work}
\label{s:Summary and future work}

This work describes the first calculation of gravitational waves, and
asymptotic radiative fluxes of energy and axial angular momentum,
produced by a spinless test mass on a generic bound geodesic of a
spinning black hole.  It represents the natural extension of the
earlier works shown in Table \ref{t:history}.  The future direction of
this work is rather straightforward: fill in the last checkmark in
Table \ref{t:history} by implementing an evolution scheme for the
constants of the geodesic orbits, and in particular for the Carter
constant.  Our anticipated path toward this goal is discussed in more
detail in Ref.~\cite{Hughes et al}.  Once completed, we will be able
to compute generic inspiral waveforms which will facilitate the
initial detection of EMRIs.  Completing this program will require
implementing Mino's scheme for evolving the Carter constant into
equations which can be entered directly into a code \cite{Mino, Hughes
et al, Drasco Flanagan Hughes, sthn2005}.  It will also require a
variety of more computational tasks, such as developing a dynamical
scheme for artificially zeroing badly behaved modes as described in
Secs.~\ref{ss:Truncation} and \ref{ss:Modes}. We are optimistic about
completing this program since most of the remaining work is relatively
straightforward in the sense that there are no known potentially
overwhelming hurdles.

\begin{acknowledgments}
We thank Marc Favata, \'Etienne Racine, Saul Teukolsky, and especially
\'Eanna Flanagan for helpful discussions. 
We also thank Enrico Barausse for identifying several typographical errors in 
equations contained in the appendices of an earlier version of this manuscript.  
We are particularly 
grateful to Kostas Glampedakis and Daniel Kennefick for providing data
that allowed us to compare results with their code \cite{equatorial},
and to Norichika Sago for comparing some of our generic fluxes with 
analytical post-Newtonian approximations \cite{Sago in prep}.
We also thank Eric Poisson, Adam Pound, and Bernhard Nickel for
providing an advance draft of Ref.\ {\cite{ppn05}} to us, and Eric
Poisson for extensive discussion regarding the importance of the
conservative self force.  S.D.~would like to thank Takahiro Tanaka,
Kyoto University, and the Yukawa Institute for Theoretical Physics,
for hospitality during the final stages of editing this manuscript.
This work was supported at Cornell by NSF Grants PHY-0140209 and
PHY-0457200, and the NASA/New York Space Grant Consortium, and at MIT
by NASA Grant NAGW-12906 and NSF Grant PHY-0244424.  
Part of this research was carried out at the Jet Propulsion
Laboratory, California Institute of Technology, under a contract with
the National Aeronautics and Space Administration and funded through the 
internal Research and Technology Development program.
\end{acknowledgments}

\appendix

\section{Geodesic details}
\label{s:Geodesic details}

The functions $V_t$, $V_r$, $V_\theta$, and $V_\phi$ appearing in the
geodesic equations (\ref{geodesics}) are given by \cite{MTW,Carter}:
\begin{subequations} \label{detailed geodesics}
\begin{align}
V_t(r,\theta)
  &= E \left( \frac{\varpi^4}{\Delta} - a^2\sin^2\theta \right)
     + aL_z \left( 1 - \frac{\varpi^2}{\Delta} \right),&
\label{tdot} \\
V_r(r)
  &= \left( E\varpi^2 - a L_z \right)^2
  - \Delta\left[\mu^2 r^2 + (L_z - a E)^2 + Q\right],&
\label{rdot}\\
V_\theta(\theta) &= Q - L_z^2 \cot^2\theta - a^2(\mu^2 - E^2)\cos^2\theta,&
\label{thetadot}\\
V_\phi(r,\theta)
  &= L_z \csc^2\theta + aE\left(\frac{\varpi^2}{\Delta} - 1\right) - \frac{a^2L_z}{\Delta},&
\label{phidot}
\end{align}
\end{subequations}
where we have defined
\begin{equation}
\varpi^2 = r^2 + a^2.
\end{equation}

We now show some explicit relations which are used to 
evaluate functions of the geodesics. As indicated below, these 
relations were first derived either by Schmidt \cite{Schmidt} 
or in Ref.~\cite{Drasco Hughes}.  We reproduce them here for the sake 
of completeness.

When evaluated in order, the following equations determine the energy
$E = \mu\tilde E$, angular momentum $L_z = \mu M \tilde L_z$, and
Carter constant $Q = (\mu M)^2 \tilde Q$ for an orbit of a given
minimum radius $\tilde r_1 = r_{\min}/M$, maximum radius $\tilde r_2 =
r_{\max}/M$, and orbital inclination $\theta_\text{inc} = \pi/2 -
\theta_{\min}$:
\begin{eqnarray}
\tilde E^2 &=& 
\frac{\kappa \rho + 2\varepsilon\sigma - 2D\sqrt{\sigma(\sigma\varepsilon^2 + \rho\varepsilon\kappa - \eta\kappa^2)}}
     {\rho^2 + 4\eta\sigma},
\\
\tilde L_z &=& -\frac{g_1\tilde E}{h_1} 
              + D \sqrt{  \frac{g_1^2\tilde E^2}{h_1^2} + \frac{f_1 \tilde E^2 - d_1}{h_1}}.
\\
\tilde Q &=& z_- \left( \beta + \frac{\tilde L_z^2}{1 - z_-}\right).
\end{eqnarray}
These were derived by Schmidt in Appendix B of Ref.~\cite{Schmidt}. 
Here the constant $D = \sgn L_z$, so that $D = 1$ for a prograde orbit, and $D = -1$ for a retrograde orbit.
The other constants in these equations are defined in terms of the constants\footnote{Note 
that Schmidt defines $z_-$ as $\cos\theta_{\min}$ \cite{Schmidt}.}
\begin{align}
&\tilde a = a/M,&
&z_- = \cos^2\theta_{\min},&
&\beta = \tilde a^2 (1 - \tilde E^2),&
\end{align}
the functions 
\begin{eqnarray}
\tilde \Delta(\tilde r) &=& \tilde r^2 - 2\tilde r + \tilde a^2,
\\
d(\tilde r) &=& \tilde \Delta (\tilde r^2 + z_- \tilde a^2),
\\
f(\tilde r) &=& \tilde r^4 + \tilde a^2 \left[ \tilde r(r+2) + z_- \tilde \Delta \right],
\\
g(\tilde r) &=& 2\tilde a \tilde r,
\\
h(\tilde r) &=& \tilde r(\tilde r - 2) + \frac{z_-\tilde \Delta}{1 - z_-},
\end{eqnarray}
and the determinants
\begin{eqnarray}
\kappa &=& d_1h_2 - d_2h_1,
\\
\varepsilon &=& d_1g_2 - d_2g_1,
\\
\rho &=& f_1h_2 - f_2h_1,
\\
\eta &=& f_1g_2 - f_2g_1,
\\
\sigma &=& g_1h_2 - g_2h_1.
\end{eqnarray}
A subscript $1,2$ means the function is to be evaluated at $\tilde
r_{1,2}$.

Schmidt also derives the following expressions for the coordinate-time 
frequencies:
\begin{align}
M \Omega_r &= \frac{\pi p K(k)}{(1 - e^2)\left[ (W + \tilde a^2 z_+ \tilde E X)K(k) - \tilde a^2 z_+ \tilde E X E(k) \right]},&
\label{Momega_r} \\
M \Omega_\theta &= \frac{\pi \beta X\sqrt{z_+}}{2\left[ (W + \tilde a^2 z_+ \tilde E X)K(k) - \tilde a^2 z_+ \tilde E X E(k) \right]},&
\label{Momega_theta}\\
M \Omega_\phi &= \frac{(Z-\tilde L_z X)K(k) + \tilde L_z X \Pi(\pi/2,z_-,k)}{(W + \tilde a^2 z_+ \tilde E X)K(k) - \tilde a^2 z_+ \tilde E X E(k)}.&
\label{Momega_phi}
\end{align}
Here $k = \sqrt{z_- / z_+}$,
\begin{equation}
z_+ = \frac{\tilde L_z^2 + \tilde Q + \beta + \sqrt{(\tilde L_z^2 + \tilde Q + \beta)^2 - 4\beta Q}}{2\beta},
\end{equation}
$K(k)$ is the complete elliptic integral of the first kind, $E(k)$ is the 
complete elliptic integral of the second kind, and $\Pi(\phi,n,k)$ is the 
Legendre elliptic integral of the third kind \cite{NR}:
\begin{align}
K(k) &= \int_0^{\pi/2} \frac{d\theta}{\sqrt{1 - k^2 \sin^2 \theta}},&
\\
E(k) &= \int_{0}^{\pi/2} d\theta \sqrt{1 - k^2 \sin^2\theta},&
\\
\Pi(\phi,n,k) &=  \int_0^{\phi} \frac{d\theta}{(1 - n \sin^2 \theta)\sqrt{1 - k^2 \sin^2 \theta}}.&
\end{align}
The remaining quantities in Eqs.~(\ref{Momega_r})-(\ref{Momega_phi})
are defined by the following integrals:
\begin{align}
W &= \int_0^{\pi} \frac{p^2 F(\chi)d\chi}{(1+e\cos\chi)^2H(\chi)\sqrt{J(\chi)}}
\\
X &= \int_0^{\pi}\frac{d\chi}{\sqrt{J(\chi)}},&
\\
Y &= \int_0^{\pi} \frac{p^2d\chi}{(1+e\cos\chi)^2\sqrt{J(\chi)}},&
\\
Z &= \int_0^{\pi} \frac{G(\chi)d\chi}{H(\chi)\sqrt{J(\chi)}},&
\end{align}
where the functions $F,G,H,J$ are given by:
\begin{align}
F(\chi) &= \tilde E + \frac{\tilde a^2 \tilde E}{p^2}(1+e\cos\chi)^2 &
  \nonumber \\
  &- \frac{2\tilde a(\tilde L_z - \tilde a \tilde E)}{p^3}(1+e\cos\chi)^3,&
\\
G(\chi) &= \tilde L_z - \frac{2}{p}(\tilde L_z - \tilde a \tilde E)(1+e\cos\chi),&
\\
H(\chi) &= 1 - \frac{2}{p}(1+e\cos\chi) + \frac{\tilde a^2}{p^2}(1+e\cos\chi)^2,&
\\
J(\chi) &= (1-\tilde E^2)(1 - e^2)&
  \nonumber\\
  &+2\left(1 - \tilde E^2 -\frac{1-e^2}{p}\right)(1+e\cos\chi)&
  \nonumber\\
  &+\left\{ (1-\tilde E^2)\frac{3+e^2}{1-e^2} - \frac{4}{p} + \frac{1-e^2}{p^2} \right.&
  \nonumber \\
  &\left. \times \left[ \tilde a^2(1 - \tilde E^2) + \tilde L_z^2 + \tilde Q \right] \right\}(1+e\cos\chi)^2&
\end{align}
The Mino time frequencies are then found using
$\Upsilon_{\phi,\theta,r} = \Gamma \omega_{\phi,\theta,r}$, and
$\Upsilon_\theta = \pi \sqrt{\beta z_+} / [2K(k)]$ (from
Ref.~\cite{Drasco Hughes}).

The function $w_r(\psi)$ and its derivative were derived in the
Appendix of Ref.~\cite{Drasco Hughes}; however, there we did not write
things explicitly in terms of Schmidt's $J$-function.  The results
are:
\begin{eqnarray}
w_r(\psi) &=& \frac{1 - e^2}{p} \int_0^{\psi}
\frac{\Upsilon_r\,d\psi'}{\sqrt{J(\psi')}},
\\
\frac{dw_r}{d\psi}(\psi) &=& \frac{1 - e^2}{p}\frac{\Upsilon_r}{\sqrt{J(\psi)}}.
\end{eqnarray}
The function $w_\theta(\chi)$ and its derivative was derived in the 
Appendix of Ref.~\cite{Drasco Hughes}.  The results are
\begin{align}
w_\theta(\chi) &= \left\{
\begin{array}{lc}
 \Upsilon_\theta \lambda_0(\chi)                &  0 \le \chi \le \pi/2
\\
 \pi - \Upsilon_\theta \lambda_0(\pi - \chi)    &  \pi/2 \le \chi \le \pi
\\
 \pi + \Upsilon_\theta \lambda_0(\chi - \pi)    &  \pi \le \chi \le 3\pi/2
\\
 2\pi - \Upsilon_\theta \lambda_0(2\pi - \chi)  &  3\pi/2 \le \chi \le 2\pi
\end{array}
\right. ,&
\\
\frac{dw_\theta}{d\chi}(\chi) &= \frac{\pi}{2K(k)}\frac{1}{1-k^2 \cos^2\chi},&
\end{align}
where 
\begin{equation}
\lambda_0(\chi) = \frac{1}{\sqrt{\beta z_+}} \left[ K(k) - \mathcal{F}(\pi/2-\chi,k)\right],
\end{equation}
and $\mathcal{F}(\phi,k)$ is the incomplete elliptic integral of the
first kind \cite{NR}:
\begin{equation}
\mathcal{F}(\phi,k) = \int_0^\phi \frac{d\theta}{\sqrt{1 - k^2 \sin^2\theta}}.
\end{equation}

\section{Perturbation details}
\label{s:Perturbation details}
The differential operators in the master equation (\ref{master}) are
given by
\begin{subequations}
\begin{align}
\widehat{U}_{t \phi r}(r) = 
&\frac{(r^2 + a^2)^2}{\Delta}          \frac{\partial^2}{\partial t^2}
+ 4\left[ \frac{M(r^2-a^2)}{\Delta} - r\right] \frac{\partial}{\partial t}& 
\nonumber \\
&+ \frac{a^2}{\Delta}    \frac{\partial^2}{\partial \phi^2}
+ \frac{4a}{\Delta}(r-M) \frac{\partial }{\partial \phi}& 
\nonumber \\
&+ \frac{4Mar}{\Delta} \frac{\partial^2 }{\partial t \partial \phi}
- \Delta^{2}           \frac{\partial}{\partial r} \left( \frac{1}{\Delta} \frac{\partial}{\partial r}\right),&
\label{Utphir} \\
\widehat{V}_{t \phi \theta}(\theta) = 
&- a^2 \sin^2\theta \frac{\partial^2}{\partial t^2}
- 4ia \cos\theta    \frac{\partial}{\partial t}&
\nonumber \\
&- \frac{1}{\sin^2 \theta}           \frac{\partial^2 }{\partial \phi^2}
+ 4\frac{i \cos\theta}{\sin^2\theta} \frac{\partial}{\partial \phi}&
\nonumber \\
&- \frac{1}{\sin \theta} \frac{\partial}{\partial \theta} \left( \sin \theta \frac{\partial \psi}{\partial \theta}\right)
+ 4\cot^2\theta + 2.&
\label{Vtphitheta}
\end{align}
\end{subequations}

In Boyer-Lindquist coordinates, the Kinnersly tetrad vectors are given by \cite{circular,Chandra}
\begin{subequations}\label{K vectors}
\begin{align}
\vec l &= \frac{\varpi^2}{\Delta} \vec \partial_t 
       + \vec \partial_r 
       + \frac{a}{\Delta} \vec \partial_\phi,&
\\
\vec m &= -\frac{\rho}{\sqrt{2}} \left(ia \sin \theta \vec \partial_t 
                                 + \vec \partial_\theta 
                                 + \frac{i}{\sin \theta} \vec \partial_\phi \right),&
\\
\vec n &= \frac{\varpi^2}{2\Sigma} \vec \partial_t 
      - \frac{\Delta}{2\Sigma}\vec \partial_r
      + \frac{a}{2\Sigma} \vec \partial_\phi,&
\end{align}
\end{subequations}
and the Kinnersly tetrad one-forms are
\begin{subequations}\label{K 1forms}
\begin{align}
\widetilde l &=-\widetilde{dt} + \frac{\Sigma}{\Delta}~\widetilde{dr}+a\sin^2\theta~\widetilde{d\phi},&
\\
\widetilde m &=\frac{\rho}{\sqrt{2}}\left( ia\sin\theta~\widetilde{dt} - \Sigma~\widetilde{d\theta} - i\varpi^2 \sin\theta~\widetilde{d\phi} \right),&
\\
\widetilde n &=-\frac{\Delta}{2\Sigma}~\widetilde{dt} -\frac{1}{2}~\widetilde{dr} + \frac{a\Delta \sin^2\theta}{2\Sigma}~\widetilde{d\phi}.&
\end{align}
\end{subequations}

The $A_{abc}$ functions are given by
\footnote{Note that the code used in Refs.~\cite{circular,circularII}
had the wrong prefactor for $A_{nn0}$ (it had the factor of $1/\bar\rho$
replaced with a factor of $1/\rho$).  Also Ref.~\cite{circular}
has an incorrect expression for $A_{n\bar m 1}$ (though this term is
correct in the code for Refs.~\cite{circular,circularII})}
\begin{subequations} \label{A's}
\begin{align}
A_{nn0}
  &= -\frac{2\rho^{-3}\bar\rho^{-1}C_{nn}}{\Delta^2}
  \left(L^\dag_1L^\dag_2 S + 2 i a \rho L^\dag_2 S \sin\theta\right),&
  \label{eq:Ann0}
\\
A_{n\bar m0}
  &= -\frac{2\sqrt{2}\rho^{-3}C_{n\bar m}}{\Delta}
  \left[\left(\frac{iK}{\Delta} - \rho - \bar\rho\right)L^\dag_2 S \right.&
  \nonumber \\
  & \left. + \left( \frac{iK}{\Delta} + \rho + \bar\rho \right)
  ia(\rho - \bar\rho)S\sin\theta\right] ,&
  \label{eq:Anm0}
\\
A_{\bar m\bar m 0}
  &= S\rho^{-3}\bar\rho C_{\bar m \bar m}
  \left[\left(\frac{K}{\Delta}\right)^2
  + 2 i \rho \frac{K}{\Delta} +
  i \partial_r\left(\frac{K}{\Delta}\right)\right],&
  \label{eq:Amm0}
\\
A_{n\bar m1}
  &= -\frac{2\sqrt{2}\rho^{-3}C_{n\bar m}}{\Delta}
  \left[L^\dag_2 S + ia(\rho - \bar\rho) S \sin\theta\right] ,&
  \label{eq:Anm1}
\\
A_{\bar m\bar m1}
  &= 2S\rho^{-3}\bar\rho\,C_{\bar m\bar m}
  \left(\rho - \frac{iK}{\Delta}\right) ,&
\label{eq:Amm1}
\\
A_{\bar m\bar m2}
  &= -S\rho^{-3}\bar\rho\,C_{\bar m\bar m} ,&
  \label{eq:Amm2}
\end{align}
\end{subequations}
where we have used the shorthand notation $S = S_{lm}(\theta,\omega)$.
Note that the factor of $C_{\bar m \bar m}$ in Eq.~(\ref{eq:Amm0})
was missing in Ref.~\cite{circular} (it was not however missing from 
the numerical code associated with Ref.~\cite{circular}).


\end{document}